\documentstyle[epsfig,12pt]{article}
\newcommand{\drawsquare}[2]{\hbox{%
\rule{#2pt}{#1pt}\hskip-#2pt%  left vertical
\rule{#1pt}{#2pt}\hskip-#1pt%  lower horizontal
\rule[#1pt]{#1pt}{#2pt}}\rule[#1pt]{#2pt}{#2pt}\hskip-#2pt%upper horizontal
\rule{#2pt}{#1pt}}% right vertical

\newcommand{\Yfund}{\raisebox{-.5pt}{\drawsquare{6.5}{0.4}}}%  fund
\newcommand{\Ysymm}{\Yfund\hskip-0.4pt%
                    \Yfund}%  symmetric second rank
\def\symm{\Ysymm}

\def\drawbox#1#2{\hrule height#2pt
        \hbox{\vrule width#2pt height#1pt \kern#1pt
              \vrule width#2pt}
              \hrule height#2pt}

\def\Asym#1#2{\vcenter{\vbox{\drawbox{#1}{#2}
              \kern-#2pt       % line up boxes
              \drawbox{#1}{#2}}}}

\def\asymm{\Asym{6.4}{0.3}}

\def\beqn{\begin{eqnarray}}
\def\eeqn{\end{eqnarray}}

\def\beq{\begin{equation}}
\def\eeq{\end{equation}}
\def\ba{\beq\new\begin{array}{c}}
\def\ea{\end{array}\eeq}

\newcommand{\nfour}{${\cal N}=4\;$}
\newcommand{\ntwo}{${\cal N}=2\;$}

\newcommand{\vp}{\varphi}
\newcommand{\pt}{\partial}

\begin{document}

\setcounter{footnote}0
%\begin{titlepage}
%\renewcommand{\thefootnote}{\fnsymbol{footnote}}

\vfill

%%%%%%%%%%%%%%%%%%%%%%%%%%%%%%%%
\begin{titlepage}

\begin{flushright}
FTPI-MINN-04/41, UMN-TH-2327/04\\
NSF-KITP-04-125, ITEP-TH-46/04\\
IHES-52/04
%%%December  7, 2004
%December  13, 2004
\end{flushright}

\begin{center}

{ \Large \bf   Non-Abelian Meissner Effect in Yang--Mills Theories
at Weak Coupling}
\end{center}

\begin{center}
 { \bf A.~Gorsky\,$^{a,b,c}$,
 \bf    M.~Shifman$^{b,c}$ and \bf A.~Yung$^{a,b,d}$}
\end {center}
\vspace{0.3cm}
\begin{center}
$^a${\it Institute of Theoretical and Experimental Physics, Moscow
117259, Russia}\\
$^b${\it  William I. Fine Theoretical Physics Institute,
University of Minnesota,
Minneapolis, MN 55455, USA}\\
$^{c}${\it  Kavli Institute for Theoretical Physics,
 UCSB, Santa Barbara, CA 93106, USA}\\
$^{d}${\it Petersburg Nuclear Physics Institute, Gatchina, St. Petersburg 188300, Russia}
\end{center}

\vspace*{.45cm}
\begin{center}
{\large\bf Abstract}
\end{center}
\vspace*{.05cm}
We present a weak-coupling 
Yang--Mills model supporting non-Abelian 
magnetic flux tubes and non-Abelian confined magnetic monopoles.
In the dual description the magnetic flux tubes are prototypes
of the QCD strings. Dualizing the confined magnetic monopoles
we get gluelumps which convert a ``QCD string" in the excited state
to that in the ground state. Introducing a mass parameter $m$
we discover a phase transition between the Abelian and non-Abelian confinement 
at a critical value $m=m_*\sim \Lambda$. Underlying dynamics are 
governed by a $Z_N$ symmetry inherent to the model under consideration.
At $m>m_*$ the $Z_N$ symmetry is spontaneously broken,
resulting in $N$ degenerate $Z_N$ (Abelian) strings. At $m<m_*$ the $Z_N$ 
symmetry is restored, the degeneracy is lifted, and the strings 
become non-Abelian. We calculate tensions of the non-Abelian
strings, as well as the decay rates of the metastable strings,
at $N\gg 1$.

\end{titlepage}

\tableofcontents

\newpage
\section{Introduction}
\label{introd}

Ever since  't Hooft \cite{thooft}
and Mandelstam \cite{mandelstam}
put forward the hypothesis of the dual Meissner effect 
to explain color confinement in non-Abelian gauge
theories  people
were trying to find a controllable
approximation in which one could reliably demonstrate
the occurrence of the dual Meissner effect in these theories. 
A breakthrough achievement was the Seiberg-Witten
solution \cite{sw} of ${\cal N}=2$ supersymmetric 
Yang--Mills theory. They found massless monopoles and,
adding a small $({\cal N}=2)$-breaking deformation,
proved that they condense creating strings carrying 
a chromoelectric flux. It was a great success in qualitative
understanding of color confinement.

A more careful examination shows, however, that details of
the Seiberg-Witten confinement are quite different from 
those we expect in QCD-like theories. Indeed, a crucial aspect
of Ref.~\cite{sw} is  that the SU($N$)
gauge symmetry is first broken, at a high scale, down to U(1)$^{N-1}$,
which is then completely broken, at a much lower scale
where monopoles condense. Correspondingly,
the strings in the Seiberg-Witten
solution are, in fact, Abelian strings \cite{ANO}
of the Abrikosov--Nielsen--Olesen
(ANO) type which results, in turn, in confinement
whose structure does not resemble at all that of QCD. In particular, the
``hadronic" spectrum is much richer than that in QCD
\cite{DS,matt}.

Thus, the problem of obtaining the Meissner effect in a more realistic
regime in theories which are closer relatives of QCD
remains open. A limited progress in this direction
was achieved since the 1980's \cite{limited};
the advancement accelerated in recent years 
\cite{recent,Hanany,Auzzi,ShifmanYung,Tong,HananyTong,Markov}.
Our task is to combine and distill these advances
to synthesize a relatively simple non-Abelian model
exhibiting at least some features of {\em bona fide}
non-Abelian confinement in a controllable setting.

What do we know of color confinement in QCD?
At a qualitative level surprisingly much.
We know that in the Yang--Mills theory
chromoelectric flux tubes are formed
between the probe heavy quarks (more exactly,
between the quark and its antiquark),
with the fundamental tension $T_1$ proportional to the square of the
dynamical scale parameter, which does not scale with $N$
at large $N$,
$$
T_1 \sim \Lambda^2_{\rm QCD}\, .
$$
If one pulls together $N$ such flux tubes
they can annihilate. This clearly distinguishes
QCD flux tubes from the ANO strings. We know that
for $k$-strings\,\footnote{Operationally,
$k$-strings are defined as flux tubes attached to probe sources
with $k$ fundamental or $k$ antifundamental indices.} (with $k>1$) excitations 
lie very close
to the ground state. For instance, if one considers
two-index symmetric and antisymmetric sources, the corresponding
string tensions $T_{[2]}$ and
$T_{\{2\}}$ are split 
\cite{Armoni-two} by $\Lambda^2/N^2$.
The decay rate 
of the symmetric string into antisymmetric (per unit length of the string per unit time)
is  
\beq
\Gamma_{\symm\,\, \to\, \asymm}\sim \Lambda^2 \exp\left(-\gamma \, N^2 \right)\,,
\label{expdec} 
\eeq
where $\gamma$ is a positive constant of  order one.
We would like to model all the above features at weak coupling,
where all approximations made can be checked and verified.
After extensive searches we found seemingly the simplest
Yang--Mills model which does the job, at least to an extent.
Our model seems to be minimal. It is non-supersymmetric.
It supports non-Abelian 
magnetic flux tubes and non-Abelian confined magnetic monopoles
at weak coupling.
In the dual description the magnetic flux tubes are prototypes
of the QCD strings. Dualizing the confined magnetic monopoles
we get gluelumps (string-attached gluons)
 which convert a ``QCD string" in the excited state
to that in the ground state. The decay rate of the excited string
to its ground state is suppressed exponentially in $N$.

It is worth  noting that strings
in non-Abelian theories at weak coupling were found long ago 
\cite{oldstrings} --- the so-called $Z_N$ strings associated with the center of 
the  SU($N$) gauge group. However, in all these constructions the gauge flux
was always directed along a fixed vector in the Cartan subalgebra 
of SU($N$), and no
moduli which would make the flux orientation
a dynamical variable  in the group space were ever found.
Therefore, these strings are, in essence, Abelian.

Recently, non-Abelian strings were shown to emerge at weak coupling 
\cite{Auzzi,ShifmanYung,HananyTong,Markov} in \ntwo and deformed \nfour
supersymmetric gauge theories  
(similar results in three dimensions were obtained in \cite{Hanany}). 
The  main feature of  the non-Abelian strings is the
presence of orientational zero modes associated with the rotation of their
color flux in the non-Abelian gauge group, which makes such  strings 
genuinely non-Abelian. This is as good as it gets at weak coupling.

In this paper  we extend (and simplify) the class of theories in
which non-Abelian strings are supported. To this end we consider 
a ``minimal" non-supersymmetric
gauge theory with the gauge group SU($N$)$\times$U(1).
Our model is still rather far  from real-world QCD. We
believe, however,  that our non-Abelian strings capture basic features
of QCD strings to a much greater extent than the  Abelian ANO strings.

Striking similarities between four-dimensional gauge theories and 
two-dimensio\-nal sigma models were noted  long ago, in 
the 1970's and 80's.  We continue revealing
reasons lying behind these similarities: in fact,  
two-dimensional sigma models
are effective low-energy theories describing  orientational moduli on the world sheet
of non-Abelian confining strings. A particular direct relation was found
previously  in  \ntwo supersymmetric theories 
\cite{Dorey:1998yh,Dorey:1999zk,ShifmanYung,HananyTong}
where the BPS kink spectrum in two-dimensional
$CP(N-1)$ model coincides with the dyon spectrum of 
a four-dimensional gauge
theory given by the exact Seiberg-Witten solution. Pursuing this line of 
research we reveal a similar relationship
between non-supersymmetric
two- and four-dimensional theories. The physics of non-supersymmetric 
sigma models significantly 
differs from that of supersymmetric ones. We find interpretations 
of known results on non-supersymmetric $CP(N-1)$ models in terms 
of non-Abelian strings  and monopoles in four dimensions.

In particular,  in parallel
to the supersymmetric case \cite{Tong,ShifmanYung,HananyTong},
we interpret the confined monopole realizing  a junction of two
distinct  non-Abelian
strings,   as a kink in the  two-dimensional $CP(N-1)$ model.
The argument is made explicit by virtue of an extrapolation procedure
designed specifically for this purpose. Namely, 
we introduce   mass parameters
$m_A$ ($A=1,...,N_f$, and $N_f=N$ is the number of bulk flavors)
for scalar quarks in four dimensions. This lifts the orientational moduli of the string.
Now the effective world-sheet description of the string internal 
dynamics   is given by a massive $CP(N-1)$ model.
In this quasiclassical limit the matching 
between the magnetic monopoles and kinks
is rather obvious. Tending $m_A \to 0$ we extrapolate this matching to the
quantum regime.

In addition to  the four-dimensional
confinement, that ensures
that the magnetic monopoles are attached to the strings, they are also
confined in the two-dimensional sense. Namely, the monopoles stick to
anti-monopoles on the string
they are attached to, to form   meson-like configurations.
The  two-dimensional
confinement disappears if the vacuum angle 
$\theta=\pi$. Some monopoles become deconfined along the string.
Alternatively, one can say that strings become degenerate.

With non-vanishing mass terms of the type
$$
\{m_A\} \longrightarrow m  \,\,   \left\{  e^{2\pi i/N} , \,e^{4\pi i/N},\, ... ,
 \,e^{2(N-1)\pi i/N} , \, 1\right\} \,,
$$
 a discrete $Z_N$ symmetry survives in the effective 
world-sheet theory. In the domain of large $m$ 
(large compared to the scale of the $CP(N-1)$ model) we have Abelian strings
and essentially the 't Hooft-Polyakov
monopoles, while at small $m$ the strings and monopoles
we deal with  become
non-Abelian. We show that 
these two regions are separated by a phase transition (presumably,
of the second order)
which we interpret as a transition between the Abelian and non-Abelian
confinement. 
We show that in the effective $CP(N-1)$ model on the string world sheet
this phase transition is associated with the 
restoration of $Z_N$ symmetry: $Z_N$ symmetry is broken in the
Abelian confinement phase and restored in the non-Abelian confinement
phase. This is a key result 
of the present work which has an intriguing 
(albeit, rather remote) parallel with the breaking
of the $Z_N$ symmetry  at the confinement/deconfinement phase transition
 found in lattice QCD at non-zero temperature.

Next, we consider some special features of  the simplest  
SU(2)$\times$U(1) case.  In particular, we
discuss the vacuum angle dependence. $CP(1)$ model is known to 
become conformal at $\theta=\pi$, including massless monopoles/kinks
at $\theta=\pi$.

Finally, we focus on the problem of the multiplicity of the hadron 
spectrum in the general SU($N$)$\times$U(1) case. 
As was already mentioned, the Abelian confinement generates too many
hadron states as compared to QCD-based  expectations  \cite{DS,matt,VY}.
In our model this regime occurs at large $m_A$. 

\section{In search of non-Abelian strings and monopoles}
\label{search}

A reference model which we suggest for
consideration is quite simple.  The gauge group of the model
is SU($N)\times$U(1). Besides SU($N$) and U(1)
gauge bosons   
the model contains $N$ scalar fields charged with respect to
U(1) which form $N$ fundamental representations of SU($N$).
It is convenient to write these fields in the form of 
$N\times N$ matrix $\Phi =\{\varphi^{kA}\}$
where $k$ is the SU($N$) gauge index while $A$ is the flavor
index, 
\beq
\Phi =\left(
\begin{array}{cccc}
\varphi^{11} & \varphi^{12}& ... & \varphi^{1N}\\[2mm]
\varphi^{21} & \varphi^{22}& ... & \varphi^{2N}\\[2mm]
...&...&...&...\\[2mm]
\varphi^{N1} & \varphi^{N2}& ... & \varphi^{NN}
\end{array}
\right)\,.
\label{phima}
\eeq
Sometimes we will refer to $\varphi$'s as to scalar quarks, or just quarks.
The action of the model has the form\,\footnote{Here and below we use a formally  
Euclidean notation, e.g.
$F_{\mu\nu}^2 = 2F_{0i}^2 + F_{ij}^2$,
$\, (\partial_\mu a)^2 = (\partial_0 a)^2 +(\partial_i a)^2$, etc.
This is appropriate since we are  going to study static (time-independent)
field configurations, and $A_0 =0$. Then the Euclidean action is
nothing but the energy functional.}
\beqn
S &=& \int {\rm d}^4x\left\{\frac1{4g_2^2}
\left(F^{a}_{\mu\nu}\right)^{2}
+ \frac1{4g_1^2}\left(F_{\mu\nu}\right)^{2}
 \right.
 \nonumber\\[3mm]
&+&
 {\rm Tr}\, (\nabla_\mu \Phi)^\dagger \,(\nabla^\mu \Phi )
+\frac{g^2_2}{2}\left[{\rm Tr}\,
\left(\Phi^\dagger T^a \Phi\right)\right]^2
 +
 \frac{g^2_1}{8}\left[ {\rm Tr}\,
\left( \Phi^\dagger \Phi \right)- N\xi \right]^2 
 \nonumber\\[3mm]
 &+&\left.
 \frac{i\,\theta}{32\,\pi^2} \, F_{\mu\nu}^a \tilde{F}^{a\,\mu\nu}
 \right\}\,,
\label{redqed}
\eeqn
where $T^a$ stands for the generator of the gauge SU($N$),
\beq
\nabla_\mu \, \Phi \equiv  \left( \partial_\mu -\frac{i}{\sqrt{ 2N}}\; A_{\mu}
-i A^{a}_{\mu}\, T^a\right)\Phi\, ,
\label{dcde}
\eeq
(the global flavor SU($N$) transformations then act on $\Phi$ from the 
right), and $\theta$ is the vacuum angle. The action (\ref{redqed}) in 
fact represents a truncated bosonic sector of the \ntwo model. The last 
term in the second line
forces $\Phi$ to develop a vacuum expectation value (VEV) while the 
last but one term
forces the VEV to be diagonal,
\beq
\Phi_{\rm vac} = \sqrt\xi\,{\rm diag}\, \{1,1,...,1\}\,.
\label{diagphi}
\eeq

In this paper we assume the parameter $\xi$ to be large,\footnote{The reader may
recognize $\xi$ as a descendant of the Fayet--Iliopoulos parameter.}
\beq
\sqrt{\xi}\gg \Lambda_4,
\label{weakcoupling} 
\eeq
where $\Lambda_4$ is the scale of the four-dimensional theory (\ref{redqed}).
This ensures the weak coupling regime as both couplings $g^2_1$ and $g^2_2$
are frozen at a large scale.

The  vacuum field (\ref{diagphi}) results in  the spontaneous
breaking of both gauge and flavor SU($N$)'s.
A diagonal global SU($N$) survives, however,
namely
\beq
{\rm U}(N)_{\rm gauge}\times {\rm SU}(N)_{\rm flavor}
\to {\rm SU}(N)_{\rm diag}\,.
\eeq
Thus, color-flavor locking takes place in the vacuum.
A version of this  scheme of symmetry breaking was suggested
long ago \cite{BarH}.

Now, let us briefly review string solutions in this model.
Since it includes a spontaneously broken gauge U(1),
the model supports
conventional ANO strings \cite{ANO}
in which one can discard the SU($N$)$_{\rm gauge}$ part 
of the action.
The topological stability of the ANO string is due to the fact that
$\pi_1({\rm U(1)}) = Z$. These are not the strings we are interested in.
At first sight the triviality of the homotopy group, $\pi_1 ({\rm SU}(N)) =0$, 
implies that there are no other topologically stable strings.
This impression is false. One can
combine the $Z_N$ center of SU($N$) with the elements $\exp (2\pi i k/N)\in$U(1) 
to get a topologically stable string solution
possessing both windings, in SU($N$) and U(1). In other words,
\beq
\pi_1 \left({\rm SU}(N)\times {\rm U}(1)/ Z_N
\right)\neq 0\,.
\eeq
It is easy to see that this nontrivial topology amounts to winding
of just one element of $\Phi_{\rm vac}$, say, $\vp^{11}$, or
$\vp^{22}$, etc, for instance,\footnote{As explained below,
$\alpha$ is the angle of
the coordinate  $\vec{x}_\perp$ in the perpendicular plane.}
\beq
\Phi_{\rm string} = \sqrt{\xi}\,{\rm diag} ( 1,1, ... ,e^{i\alpha (x) })\,,
\quad x\to\infty \,.
\label{ansa}
\eeq
Such strings can be called elementary;
their tension is $1/N$-th of that of the ANO string.
The ANO string can be viewed as a bound state of 
$N$ elementary strings.

More concretely,  the $Z_N$ string solution 
(a progenitor of the non-Abelian string) can be written as
follows \cite{Auzzi}:
\beqn
\Phi &=&
\left(
\begin{array}{cccc}
\phi(r) & 0& ... & 0\\[2mm]
...&...&...&...\\[2mm]
0& ... & \phi(r)&  0\\[2mm]
0 & 0& ... & e^{i\alpha}\phi_{N}(r)
\end{array}
\right) ,
\nonumber\\[5mm]
A^{{\rm SU}(N)}_i &=&
\frac1N\left(
\begin{array}{cccc}
1 & ... & 0 & 0\\[2mm]
...&...&...&...\\[2mm]
0&  ... & 1 & 0\\[2mm]
0 & 0& ... & -(N-1)
\end{array}
\right)\, \left( \pt_i \alpha \right) \left[ -1+f_{NA}(r)\right] ,
\nonumber\\[5mm]
A^{{\rm U}(1)}_i &=& \frac{1}{N}\, 
\left( \pt_i \alpha \right)\left[1-f(r)\right] ,\qquad A^{{\rm U}(1)}_0=
A^{{\rm SU}(N)}_0 =0\,,
\label{znstr}
\eeqn
where $i=1,2$ labels coordinates in the plane orthogonal to the string
axis and $r$ and $\alpha$ are the polar coordinates in this plane. The profile
functions $\phi(r)$ and  $\phi_N(r)$ determine the profiles of the scalar fields,
while $f_{NA}(r)$ and $f(r)$ determine the SU($N$) and U(1) fields of the 
string solutions, respectively. These functions satisfy the following 
rather obvious boundary conditions:
\beqn
&& \phi_{N}(0)=0,
\nonumber\\[2mm]
&& f_{NA}(0)=1,\;\;\;f(0)=1\,,
\label{bc0}
\eeqn
at $r=0$, and 
\beqn
&& \phi_{N}(\infty)=\sqrt{\xi},\;\;\;\phi(\infty)=\sqrt{\xi}\,,
\nonumber\\[2mm]
&& f_{NA}(\infty)=0,\;\;\;\; \; f(\infty) = 0
\label{bcinfty}
\eeqn
at $r=\infty$.
Because our model is equivalent, in fact, to  a bosonic reduction
 of the \ntwo supersymmetric theory,
these profile functions satisfy the first-order differential equations
obtained in \cite{MY}, namely,
\beqn
&&
r\frac{d}{{d}r}\,\phi_N (r)- \frac1N\left( f(r)
+  (N-1)f_{NA}(r) \right)\phi_N (r) = 0\, ,
\nonumber\\[4mm]
&&
r\frac{d}{{ d}r}\,\phi (r)- \frac1N\left(f(r)
-  f_{NA}(r)\right)\phi (r) = 0\, ,
\nonumber\\[4mm]
&&
-\frac1r\,\frac{ d}{{ d}r} f(r)+\frac{g^2_1 N}{4}\,
\left[(N-1)\phi(r)^2 +\phi_N(r)^2-N\xi\right] = 0\, ,
\nonumber\\[4mm]
&&
-\frac1r\,\frac{d}{{ d}r} f_{NA}(r)+\frac{g^2_2}{2}\,
\left[\phi_N(r)^2 -\phi_2(r)^2\right]  = 0\, .
\label{foe}
\eeqn
These equations  can be solved numerically. Clearly, the solutions
to the first-order equations automatically satisfy the second-order equations 
of motion. Quantum corrections destroy  fine-tuning of the coupling constants in
(\ref{redqed}). If one is interested in  calculation of the quantum-corrected 
profile functions one has to solve the second-order equations of motion
instead of (\ref{foe}).

The tension of this elementary string is 
\beq
T_1=2\pi\,\xi\, .
\label{ten}
\eeq
As soon as our theory is not supersymmetric and the string is not BPS
there are corrections to this result which are small 
and uninteresting provided the coupling 
constants $g^2_1$ and $g^2_2$ are small. Note that the tension of 
the ANO
string is 
\beq
T_{\rm ANO}=2\pi\,N\,\xi
\label{tenANO}
\eeq
in our normalization.

The elementary strings are {\em bona fide} non-Abelian.
This means that, besides trivial translational
moduli, they give rise to moduli corresponding to spontaneous
breaking of a non-Abelian symmetry. Indeed, while the ``flat"
vacuum is SU($N$)$_{\rm diag}$ symmetric, the solution (\ref{znstr})
breaks this symmetry 
down\,\footnote{At $N=2$ the string solution breaks
SU(2) down to U(1).} to U(1)$\times$SU$(N-1)$ (at $N>2$).
This means that the world-sheet (two-dimensional) theory of 
the elementary string moduli
is the SU($N$)/(U(1)$\times$ SU($N-1$)) sigma model.
This is also known as $CP(N-1)$ model.

To obtain the non-Abelian string solution from the $Z_N$ string 
(\ref{znstr}) we apply the diagonal color-flavor rotation  preserving
the vacuum (\ref{diagphi}). To this end
it is convenient to pass to the singular gauge where the scalar fields have
no winding at infinity, while the string flux comes from the vicinity of  
the origin. In this gauge we have
\beqn
\Phi &=&
U\left(
\begin{array}{cccc}
\phi(r) & 0& ... & 0\\[2mm]
...&...&...&...\\[2mm]
0& ... & \phi(r)&  0\\[2mm]
0 & 0& ... & \phi_{N}(r)
\end{array}
\right)U^{-1}\, ,
\nonumber\\[5mm]
A^{{\rm SU}(N)}_i &=&
\frac{1}{N} \,U\left(
\begin{array}{cccc}
1 & ... & 0 & 0\\[2mm]
...&...&...&...\\[2mm]
0&  ... & 1 & 0\\[2mm]
0 & 0& ... & -(N-1)
\end{array}
\right)U^{-1}\, \left( \pt_i \alpha\right)  f_{NA}(r)\, ,
\nonumber\\[5mm]
A^{{\rm U}(1)}_i &=& -\frac{1}{N}\, 
\left( \pt_i \alpha\right)   f(r)\, , \qquad A^{{\rm U}(1)}_0=
A^{{\rm SU}(N)}_0=0\,,
\label{nastr}
\eeqn
where $U$ is a matrix $\in {\rm SU}(N)$. This matrix parametrizes 
orientational zero modes of the string associated with flux rotation  
in  SU($N$). The presence of these modes makes the string genuinely
non-Abelian. Since the diagonal color-flavor symmetry is not
broken by the vacuum expectation values 
(VEV's) of the scalar fields 
in the bulk (color-flavor locking)
it is physical and has nothing to do
with the gauge rotations eaten by the Higgs mechanism. The orientational moduli
encoded in the matrix $U$ are {\it not} gauge artifacts. The orientational
zero modes of a non-Abelian string were first 
observed in \cite{Hanany, Auzzi}.

\section{The world-sheet theory for the elementary string moduli}

In this section we will present derivation of  an effective low-energy
theory for the orientational moduli of the  elementary string
and then discuss underlying physics. We will closely follow Refs. 
\cite{Auzzi,ShifmanYung} where this derivation was 
carried out  for  $N=2$ which leads to the $CP(1)$ model. In the general
case,
as was already mentioned, the resulting macroscopic theory
is  a two-dimensional $CP(N-1)$ model \cite{Hanany,Auzzi,
ShifmanYung,HananyTong}. 

\subsection{Derivation of the $CP(N-1)$ model}
\label{kineticterm}

{}First, extending  the supersymmetric $CP(1)$ derivation of   
Refs.~\cite{Auzzi,ShifmanYung},  we  will  derive  the effective low-energy
theory for the  moduli  residing in the matrix $U$ in the problem at hand.
As is clear from the string solution (\ref{nastr}),   not each element of
the matrix $U$ will give rise to a modulus. The SU($N-1) \times$U(1) subgroup 
remains unbroken 
by the string solution under consideration; therefore,  as  was already mentioned,
the moduli space is 
\beq
\frac{{\rm SU}(N)}{{\rm SU}(N-1)\times {\rm U}(1)}\sim CP(N-1)\,.
\label{modulispace}
\eeq
Keeping this in mind we parametrize the matrices entering Eq.~(\ref{nastr})
as follows:
\beq
\frac1N\left\{
U\left(
\begin{array}{cccc}
1 & ... & 0 & 0\\[2mm]
...&...&...&...\\[2mm]
0&  ... & 1 & 0\\[2mm]
0 & 0& ... & -(N-1)
\end{array}
\right)U^{-1}
\right\}^l_p=-n^l n_p^* +\frac1N \delta^l_p\,\, ,
\label{n}
\eeq
where $n^l$ is a complex vector
 in the fundamental representation of SU($N$), and
 $$
 n^*_l n^l =1\,,
 $$ 
($l,p=1, ..., N$ are color indices).
As we will show below, one U(1) phase will be gauged in the effective
sigma model. This gives the correct number of degrees of freedom,
namely, $2(N-1)$. 

With this parametrization the string solution (\ref{nastr}) can be
rewritten   as
\beqn 
\Phi &=& \frac1N[(N-1)\phi +\phi_N] -(\phi-\phi_N)\left( n\,\cdot n^*-\frac1N\right) ,
\nonumber\\[3mm]
A^{{\rm SU}(N)}_i &=& \left( n\,\cdot n^*-\frac{1}{N}\right) \varepsilon_{ij}\, 
\frac{x_i}{r^2}
\,
f_{NA}(r) \,,
\nonumber\\[3mm]
A^{{\rm U}(1)}_i &=& \frac1N
\varepsilon_{ij}\, \frac{x_i}{r^2} \, f(r) \, ,
\label{str}
\eeqn
where for brevity we suppress all SU$(N)$  indices. The notation is
self-evident.

Assume  that the orientational moduli
are slowly-varying functions of the string world-sheet coordinates
$x_{\alpha}$, $\alpha=0,3$. Then the moduli $n^l$ become fields of a 
(1+1)-dimensional sigma model on the world sheet. Since 
$n^l$ parametrize the string zero modes,
there is no potential term in this sigma model. 

To obtain the kinetic term  we substitute our solution
(\ref{str}), which depends on the
moduli $ n^l$,
in the action (\ref{redqed}), assuming  that
the fields acquire a dependence on the coordinates $x_{\alpha}$ via 
$n^l(x_{\alpha})$.
In doing so we immediately observe that we have to modify the solution
including in it the $\alpha=0,3$ components of the gauge potential
which are no more vanishing. In the $CP(1)$ case, as was  shown in 
\cite{ShifmanYung},  the potential $A_{\alpha}$ must be orthogonal 
(in the SU$(N)$  space)  to the matrix (\ref{n}) as well as to its 
derivatives with respect to
$x_{\alpha}$. Generalization of these conditions to the $CP(N-1)$ case 
leads to the 
following {\em ansatz}:
\beq
A^{{\rm SU}(N)}_{\alpha}=-i\,  \left[ \pt_{\alpha} n \,\cdot n^* -n\,\cdot 
\pt_{\alpha} n^* -2n\,
\cdot n^*(n^*\pt_{\alpha} n)
\right] \,\rho (r)\, , \qquad \alpha=0, 3\,,
\label{An}
\eeq
where we assume the contraction of the color indices inside the parentheses,
$$(n^*\pt_{\alpha} n)\equiv n^*_l\pt_{\alpha} n^l\,, $$
and introduce a new profile function $\rho (r)$. 

The function $\rho (r)$ in Eq.~(\ref{An}) is
determined  through a minimization procedure  \cite{Auzzi,ShifmanYung}
which generates $\rho$'s own equation of motion. Now we derive it.
But at first we note that
 $\rho (r)$ vanishes at infinity,
\beq
\rho (\infty)=0\,.
\label{bcfinfty}
\eeq
The boundary condition at $r=0$ will be determined shortly.

The kinetic term for $n^l$ comes from the gauge and quark kinetic terms 
in Eq.~(\ref{redqed}). Using Eqs.~(\ref{str}) and (\ref{An}) to calculate the
SU($N$)  gauge field strength we find
\beqn
F_{\alpha i}^{{\rm SU}(N)} &=& \left( \pt_{\alpha} n \,\cdot n^* 
+n\cdot \,\pt_{\alpha} n^*\right) \, \varepsilon_{ij}\,
\frac{x_j}{r^2}\,
f_{NA}
\left[1-\rho (r)\right]
\nonumber\\[3mm]
&+& 
i\left[ \pt_{\alpha} n \,\cdot n^* -n\cdot \,\pt_{\alpha} n^*
 -2n\,\cdot n^*(n^*\pt_{\alpha} n)
\right]
 \, \frac{x_i}{r}\,\, \frac{d\,\rho (r)}{dr}\, .
 \label{Fni}
\eeqn
In order to have a finite contribution  from the term
Tr$\,F_{\alpha i}^2$ in the action we have to impose the constraint
\beq
\rho (0)=1\,.
\label{bcfzero}
\eeq
Substituting the field strength (\ref{Fni}) in the action
(\ref{redqed}) and including, in addition, the  kinetic term of the quarks, 
after a rather straightforward but tedious algebra we  
arrive at
\beq
S^{(1+1)}= 2 \beta\,   \int d t\, dz \,  \left\{(\pt_{\alpha}\, n^*
\pt_{\alpha}\, n) + (n^*\pt_{\alpha}\, n)^2\right\}\,,
\label{o3}
\eeq
where the coupling constant $\beta$ is given by 
\beq
\beta=\frac{2\pi}{g^2_2}\, I \,,
\label{beta}
\eeq
and $I$ is a basic normalizing integral
\beqn
I & = &
  \int_0^{\infty}
rdr\left\{\left(\frac{d}{dr}\rho (r)\right)^2
+\frac{1}{r^2}\, f_{NA}^2\,\left(1-\rho \right)^2
\right.
\nonumber\\[4mm]
& + & 
\left.  g_2^2\left[\frac{\rho^2}{2}\left(\phi^2
 +\phi_N^2\right)+
\left(1-\rho \right)\left(\phi-\phi_N\right)^2\right]\right\}\, .
\label{I}
\eeqn

The theory in Eq.~(\ref{o3}) is in fact the two-dimensional $CP(N-1)$ model. 
To see that this is indeed the case we can eliminate
the second term in (\ref{o3})
by virtue of introduction of a non-propagating U(1) gauge field. We review
this in Sect.~\ref{Dyn}, and then discuss the underlying physics of the model.
Thus,  we obtain the  $CP(N-1)$ model as an effective low-energy
theory on the world sheet of the non-Abelian string. Its coupling $\beta$
is related to the four-dimensional coupling $g_2^2$ via 
the basic normalizing integral
(\ref{I}). This integral can be viewed as an ``action'' for the profile function $\rho$.

Varying (\ref{I}) with respect to $\rho$
one  obtains the second-order equation which 
the function $\rho$ must satisfy, namely,
\beq
-\frac{d^2}{dr^2}\, \rho -\frac1r\, \frac{d}{dr}\, \rho
-\frac{1}{r^2}\, f_{NA}^2 \left(1-\rho\right)
+
\frac{g^2_2}{2}\left(\phi_N^2+\phi^2\right)
\rho
-\frac{g_2^2}{2}\left(\phi_N-\phi\right)^2=0\, .
\label{rhoeq}
\eeq
After some algebra and extensive use of the first-order equations (\ref{foe})
one can show that the solution of (\ref{rhoeq})  is given by
\beq
\rho=1-\frac{\phi_N}{\phi}\, .
\label{rhosol}
\eeq
This solution  satisfies the boundary conditions (\ref{bcfinfty})
and  (\ref{bcfzero}).

Substituting this solution back in  the expression for the
normalizing integral  (\ref{I}) one can check that
this integral   reduces to a total derivative and is given
by the flux of the string  determined by $f_{NA}(0)=1$.
Therefore, we arrive at
\beq
I=1\,.
\label{Ieq1}
\eeq
This result can be traced back to the fact that our theory (\ref{redqed}) 
is a bosonic reduction of the
\ntwo supersymmetric theory, and the string satisfies the first-order equations
(\ref{foe}) (see \cite{ShifmanYung} for the explanation why 
(\ref{Ieq1}) should hold   for the BPS non-Abelian strings in SUSY theories).
The fact that $I=1$ was demonstrated previously for $N=2$, where  the $CP(1)$ 
model emerges.
Generally speaking, for non-BPS strings, $I$
could  be a certain  function of $N$ (see Ref.~\cite{Markov} for a 
particular example). In the problem at hand it is $N$-independent. However, 
we expect that quantum corrections
slightly  modify Eq.~(\ref{Ieq1}).

The relation between the four-dimensional and two-dimensional coupling
constants (\ref{beta}) is obtained  at the classical level. In quantum theory
both couplings run. So we have to specify a scale at which the relation
(\ref{beta}) takes place. The two-dimensional $CP(N-1)$ model 
(\ref{o3}) is
an effective low-energy theory good for the description of
internal string dynamics   at small energies,  much less than the 
inverse thickness of the string which is given by $\sqrt{\xi}$. Thus, 
$\sqrt{\xi}$ plays the role of a physical ultraviolet (UV) cutoff in (\ref{o3}). 
This is the scale at which Eq.~(\ref{beta}) holds. Below this scale, the
coupling $\beta$ runs according to its two-dimensional renormalization-group flow, 
see the next section.

\subsection{Penetration of $\theta$ from the bulk  in the world-sheet
theory}
\label{penethe}

Now let us investigate the impact of  the $\theta$ term that is present 
in our microscopic theory (\ref{redqed}).
At first sight, seemingly  it cannot produce any effect because our 
string is magnetic.
However, if one  allows for slow variations of $n^l$ with respect to $z$ 
and $t$, one immediately observes that
the electric field is generated via $A_{0,3}$ in Eq.~(\ref{An}). Substituting
$F_{\alpha i}$ from (\ref{Fni}) into the $\theta$ term in the 
action (\ref{redqed})
and taking into account   the contribution from $F_{\alpha\gamma}$ 
times $F_{ij}$
($\alpha,\gamma=0,3$ and $i,j=1,2$) we get  the topological term
in the effective $CP(N-1)$ model (\ref{o3}) in the form
\beq
S^{(1+1)}=   \int d t\, dz \,\left\{2 \beta\, 
\left[(\pt_{\alpha}\, n^*\pt_{\alpha}\, n) + (n^*\pt_{\alpha}\, n)^2\right]- 
\frac{\theta}{2\pi}\, I_{\theta}\,\, \varepsilon_{\alpha\gamma}\,
(\pt_{\alpha}\, n^*\pt_{\gamma}\, n )
\right\}\,,
\label{cpN}
\eeq
where $I_{\theta}$ is another normalizing integral given by
the formula
\beqn
I_{\theta}
&=&
- \int dr \left\{2f_{NA}(1-\rho) \, \frac{d\rho}{dr}
 +(2\rho-\rho^2) \, \frac{df}{dr}\right\}
\nonumber\\[3mm]
&=&
\int dr \frac{d}{dr}\left\{2f_{NA}\, \rho -\rho^2 \, f_{NA}\right\}.
\label{Itheta}
\eeqn
As is clearly  seen,  the integrand here reduces to a total
derivative,  and the integral is determined by the boundary conditions for
the  profile functions $\rho$ and $f_{NA}$. Substituting (\ref{bcfinfty}), 
(\ref{bcfzero})
and (\ref{bcinfty}), (\ref{bc0}) we get
\beq
I_{\theta}=1\,,
\label{Ithetaeq1}
\eeq
independently of the form of the  profile functions. This latter circumstance
 is perfectly natural
for the topological term.

The additional term (\ref{cpN}) in the $CP(N-1)$ model  
that we have just derived is the $\theta$ term in the 
standard normalization. The result (\ref{Ithetaeq1}) could have been  
expected since physics is $2\pi$-periodic with respect to 
 $\theta$ both in the 
four-dimensional microscopic gauge theory and in the effective 
two-dimensional $CP(N-1)$ model. The result (\ref{Ithetaeq1})
is not  sensitive to the presence of supersymmetry. It will hold in supersymmetric 
models as well. Note that the complexified bulk
coupling constant converts into the complexified
world-sheet coupling constant,
$$
\tau = \frac{4\pi}{g^2_2} + i\frac{\theta}{2\pi}  \to 2\beta+ i\frac{\theta}{2\pi}\,.
$$

The above derivation provides the first direct calculation 
proving the coincidence of the
$\theta$ angles in four  and two dimensions.

Let us make  a comment on this point from
the  brane perspective. Since the model under consideration is 
non-supersymmetric,  the usual brane picture corresponding
to  minimal surfaces in the external geometry is
complicated and largely unavailable at present. However,
a few statements  insensitive to details
of the brane picture can be made --- the identification
of the $\theta$ angles in the microscopic and microscopic theories above
is one of them. Indeed, in any
relevant brane picture the $\theta$ angle corresponds
to the distance between two M5 branes along the
eleventh dimension in M-theory \cite{haho}. The four-dimensional
theory is defined on the world-volume of one of these M5 branes,
while  an M2 brane stretched between M5
branes corresponds to the non-Abelian string we deal with.
It is clear that the $\theta$ angles are the same since it is just
the same geometrical parameter viewed
from two different objects: M5 and M2 branes (see also Footnote 7
in Ref.~\cite{HananyTong}).

\section{Dynamics of the world-sheet theory}
\label{Dyn}

The $CP(N-1)$ model
describing the string moduli interactions can be cast in
several equivalent representations.
The most convenient for our purposes
is a linear gauged representation
(for a review see \cite{NSVZsigma}). At large $N$ the model
was solved \cite{5,da}.

In this formulation  the Lagrangian is built from an $N$-component
complex field $n^{\ell}$ subject to the constraint
\beq
n_{\ell}^*\, n^{\ell} =1\,,
\label{lambdaco}
\eeq
The Lagrangian has the form
\beq
{\cal L} = \frac{2}{g^2}\, \left[
\left(\partial_{\alpha} + i A_\alpha\right) n^*_{\ell}
\left(\partial_\alpha - i A_\alpha \right) n^{\ell}
-\lambda \left( n^*_{\ell} n^{\ell}-1\right) 
\right]\,,
\label{one}
\eeq
where $1/g^2\equiv \beta$ and  $\lambda$ is the Lagrange multiplier
enforcing (\ref{lambdaco}). Moreover,
$A_\alpha$ is an auxiliary field which enters the Lagrangian
with no kinetic term. Eliminating $A_\alpha$ by virtue of the
equations of motion one arrives at Eq.~(\ref{cpN}).

At the quantum level the constraint (\ref{lambdaco}) is gone;
$\lambda$ becomes dynamical. Moreover,
a  kinetic term is generated 
for the auxiliary field $A_\alpha$ at the quantum level, so that    $A_\alpha$
becomes dynamical too.

As was shown above, the $\theta$ term which
can be written as
\beq
{\cal L}_\theta = \frac{\theta}{2\pi }\, \varepsilon_{\alpha\gamma} 
\partial^\alpha A^\gamma
=\frac{\theta}{2\pi }\, \varepsilon_{\alpha\gamma} \partial^\alpha \left(
n_{\ell}^*\partial^\gamma n^{\ell}
\right)
\label{two}
\eeq
appears in the world-sheet theory of the string moduli provided the same $\theta$ angle
is present in the bulk (microscopic) theory.

\begin{figure}
\epsfxsize=6cm
\centerline{\epsfbox{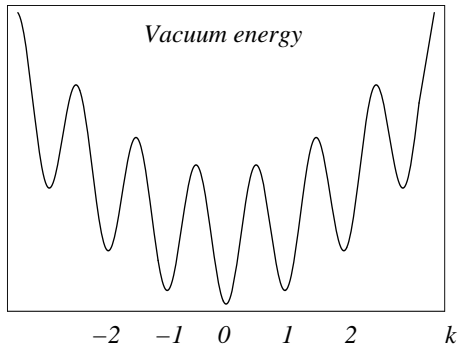}}
\caption{
The vacuum structure of 
$CP(N-1)$ model at $\theta =0$.}
\label{odin}
\end{figure}

\begin{figure}
\epsfxsize=6cm
\centerline{\epsfbox{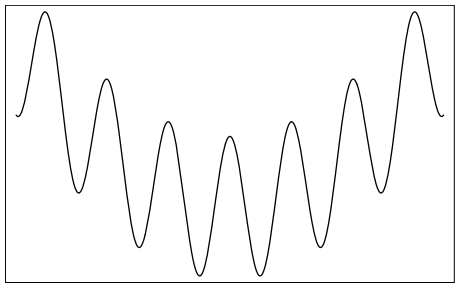}}
\caption{
The vacuum structure of 
$CP(N-1)$ model at $\theta = \pi$.}
\label{dva}
\end{figure}

Now we have to discuss the vacuum structure of the
theory (\ref{one}). Basing on a modern understanding of the
issue \cite{nvacym} (see also \cite{nvacymp}) one can say that for each $\theta$ there are
infinitely many ``vacua" that are stable in the limit $N\to \infty$.
The word ``vacua" is in the quotation marks because
only one of them presents a {\em bona fide}
global minimum; others are local minima and are metastable at
finite (but large) $N$. A schematic picture of these vacua
is given in Fig. ~\ref{odin}. All  these minima are entangled in the $\theta$ evolution.
If we vary $\theta$ continuously from $0$ to $2\pi$ the
depths of the minima ``breathe." At $\theta =\pi$ two vacua become degenerate 
(Fig.~\ref{dva}), while for larger values of $\theta$ the global minimum 
becomes local while the adjacent local minimum becomes global.
The splitting between the values of the consecutive minima
is of the order of $1/N$, while the
the probability of the false vacuum decay is proportional to
$N^{-1}\exp (-N)$, see below.

As long as the $CP(N-1)$ model plays a role of the effective theory on the
world sheet of non-Abelian string each of these ``vacua'' corresponds to
a  string in the four-dimensional bulk theory.
For each given $\theta$,  the ground state of the string
is described by the deepest vacuum of the world-sheet
theory, $CP(N-1)$. Metastable vacua of $CP(N-1)$
correspond to excited strings.

As was shown by Witten \cite{5}, 
the field $n^{\ell}$ can be viewed as a field describing kinks interpolating 
between the true vacuum and its neighbor. The multiplicity of 
such kinks is $N$ \cite{Acharya}, they form an $N$-plet. This is 
the origin of the superscript $\ell$ in $n^{\ell}$.
 
Moreover, Witten showed, by exploiting  $1/N$ 
expansion to the leading order, 
that a mass scale is dynamically generated in the model,
through dimensional transmutation,
\beq 
\Lambda^2 = M_0^2 \exp\left(-\frac{8\pi}{Ng^2}\right).
\label{scp}
\eeq
Here $M_0$ is the ultraviolet cut-off 
(for the effective theory on the string world sheet
$M_0=\sqrt{\xi}$) and $g^2=1/\beta$ is the bare
coupling constant given in Eq.~(\ref{beta}). 
The combination $Ng^2$ is nothing but
the 't Hooft constant that does not scale with $N$. 
As a result, $\Lambda$ scales as $N^0$  at large $N$.

In the leading  order, $N^0$, the kink mass $M_{n}$ is 
$\theta$-independent,
\beq
M_{n} =\Lambda\,.
\eeq
$\theta$-dependent corrections to this formula appear only at the
level $1/N^2$. 

\begin{figure}
\epsfxsize=8cm
\centerline{\epsfbox{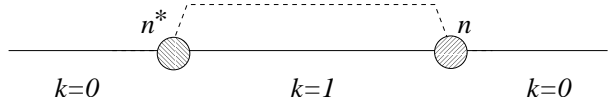}}
\caption{
Linear confinement of the $n$-$n^*$ pair. 
The solid straight line represents the string.
The dashed line shows
the vacuum energy density (normalizing ${\cal E}_0$ to zero).}
\label{tri}
\end{figure}

The kinks represented in the Lagrangian (\ref{one}) by the field $n^{\ell}$
are not asymptotic states in the $CP(N-1)$ model. In fact, 
they are confined \cite{5};
the confining potential grows linearly with distance\,\footnote{Let us note 
in passing that corrections to the leading-order result (\ref{cten}) run in 
powers of $1/N^2$ rather than $1/N$.
Indeed, as well-known, the $\theta$ dependence of the vacuum energy enters
 only through
the combination of $\theta /N$, namely ${\cal E}( \theta )  = N\Lambda^2
f( \theta /N )$ where $f $ is some function. As will be explained 
momentarily, $\Delta T = {\cal E}( \theta = 2\pi)-{\cal E}( \theta =0)$.
Moreover, ${\cal E}$, being $CP$ even, can be expanded in even powers of
$\theta$. This concludes the proof that 
$\Delta T = (12\pi\Lambda^2/N)(1 +\sum_{k=1}^\infty \, c_kN^{-2k}$).}, 
%%%end of footnote
with the tension suppressed by $1/N$. From the four-dimensional 
perspective the coefficient of the linear confinement is nothing
but the difference in tensions of two  strings:
the lightest and the next one, see below.  Therefore,  we denote it 
as $\Delta T$,
\beq
\Delta T = 12\pi\, \frac{\Lambda^2}{N}\,.
\label{cten}
\eeq
One sees that confinement  becomes
exceedingly weak at large $N$. In fact, Eq.~(\ref{cten})
refers to $\theta =0$. The standard argument that
$\theta $ dependence does not appear
at $N\to\infty$ is inapplicable to the string tension,
since the string tension itself vanishes in the large-$N$ limit.
The $\theta $ dependence can be readily established from 
a picture of the kink confinement discussed in
\cite{Markov}, see Fig.~\ref{tri},  which is complementary to that
of \cite{5}. This picture of the kink confinement is schematically 
depicted in this
figure.

Since the kink represents an interpolation between
the genuine vacuum and a false one,  the kink--anti-kink configuration
presented in Fig.~\ref{tri} shows two distinct regimes:
the genuine vacuum outside the kink--anti-kink pair
and the false one inside. As was mentioned, the string tension $\Delta T$
is  given by the difference of the vacuum energy densities,
that of the the false vacuum minus the genuine one. At large $N$,
the $k$ dependence of the energy density in the ``vacua" 
($k$ is the excitation number), as well as the $\theta$ dependence, is
well-known \cite{nvacym},
\beq
{\cal E}_k (\theta) = -\frac{6 }{\pi}\, N\, \Lambda^2
\left\{1 - \frac{1}{2}\left(\frac{2\pi k +\theta}{N}
\right)^2
\right\}
\,.
\label{split}
\eeq
At $\theta = 0$ the genuine vacuum corresponds to $k=0$,
while the first excitation to $k=-1$. At $\theta =\pi$
these two vacua are degenerate, at $\theta = 2\pi$
their roles interchange. Therefore,
\beq
\Delta T (\theta ) = 12\pi\, \frac{\Lambda^2}{N}\,\left| 1-\frac{\theta}{\pi}
\right|
\,.
\label{ctenp}
\eeq
Note that at $\theta =\pi$ the string tension vanishes
and confinement of kinks disappears.

\begin{figure}
\epsfxsize=8cm
\centerline{\epsfbox{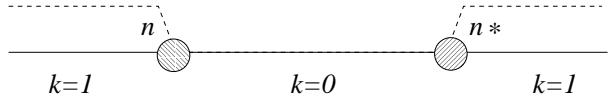}}
\caption{
Breaking of the excited string
through the $n$-$n^*$ pair creation. 
The dashed line shows
the vacuum energy density.}
\label{chetyre}
\end{figure}

\begin{figure}
\epsfxsize=8cm
\centerline{\epsfbox{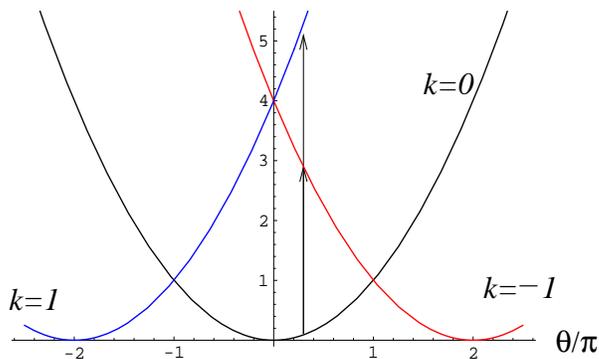}}
\caption{
The function $1 + {\cal E}_{0, \pm 1}/[ (6 N\,  \Lambda^2 )\pi^{-1}]$
in the units $\pi^2/(2N^2)$ versus $\theta /\pi$.}
\label{3k}
\end{figure}

This formula requires a comment which we hasten to make.
In fact, for each given $\theta$, there are two types of kinks
which are degenerate at $\theta =0$ but acquire a splitting at
$\theta  \neq 0$. This is clearly seen in Fig. \ref{3k}
which displays ${\cal E}_{0,\pm 1}$ for three minima:
the global one ($k=0$) and two adjacent local minima, $k= \pm 1$
(the above nomenclature refers to $ | \theta | <\pi$). 
Let us consider, say, small and positive values of $\theta$. 
Then the kink described by the field $n$
can represent two distinct interpolations:
from the ground state to the state $k=-1$
(i.e. the minimum to the left of the global minimum
in Fig.~\ref{odin}); then $$\Delta {\cal E} =
\frac{12 \pi\Lambda^2}{N} \left(1-\frac{\theta}{\pi}
\right)\,.$$  
Another possible interpolation is from the ground state to the state $k=1$
(i.e. the minimum to the right of the global minimum
in Fig.~\ref{odin}). In the latter case 
 $$\Delta {\cal E} =
\frac{12 \pi\Lambda^2}{N} \left(1+ \frac{\theta}{\pi}
\right)\,.$$  
In the first scenario the string becomes tensionless\,\footnote{Note 
that in Witten's work \cite{5} there is a misprint in Eq.~(18) and   
subsequent equations; the factor $\theta /2\pi $ should be replaced by
$\theta /\pi $. Two types of kinks correspond 
in this equation to $x >y$
and $x<y$, respectively.}, i.e. the states $k=0, -1$ degenerate,  at
$\theta = \pi$. The same consideration applies to negative values of
$\theta$.  Now it is the vacua $k=0,  1$ that become degenerate at
$\theta = -\pi$, rendering the corresponding string tensionless. 
In general, it
is sufficient to consider the interval $|\theta | \leq\pi$.

What will happen if we interchange the position of two kinks
in Fig.~\ref{tri} , as shown in Fig.~\ref{chetyre}? The excited vacuum is now outside
the kink-anti-kink pair, while the genuine one is inside.
Formally, the string tension becomes negative. In fact, the process
in Fig. 4 depicts a breaking of the excited string. As was mentioned
above, the probability of such breaking is suppressed by $\exp (-N)$.
Indeed, the master formula from Ref.~\cite{Volo} implies
that the probability of the excited string decay
(through the $n$-$n^*$ pair creation) per unit time per unit length is
\beq
\Gamma =\frac{\Delta T}{2\pi} \exp\left(-\frac{\pi M_n^2}{\Delta T}
\right) = \frac{6\Lambda^2}{N}\, e^{-N/12}
\eeq
at $\theta =0$. At $\theta \neq 0$ the suppression is even stronger.

To summarize, the  $CP(N-1)$ 
model has a fine structure of ``vacua'' which are split,
with the splitting of the order of $\Lambda^2/N$. In 
four-dimensional bulk
theory these ``vacua'' correspond to elementary non-Abelian strings.
Classically all these strings have the same tension (\ref{ten}). Due to
quantum effects in the world-sheet theory 
the degeneracy is lifted:
the elementary strings become split, with the tensions
\beq
T  = 2\pi\xi -\frac{6 }{\pi}\, N\, \Lambda^2
\left\{1 - \frac{1}{2}\left(\frac{2\pi k +\theta}{N}
\right)^2
\right\}
\,.
\label{splitten}
\eeq
Note that (i) the splitting does not appear to any finite order
in the   coupling constants; (ii) since
$\xi\gg \Lambda$,  the splitting is suppressed in both parameters, 
$\Lambda/\sqrt{\xi}$ and $1/N$.

Let us also note that the identification  of the $\theta$ terms
and topological charges in two and four
dimensions (see Sect.~\ref{penethe}) allows us to address the issue of 
$CP$ symmetry in four
dimensions at $\theta=\pi$, and confront it with
the situation  in two dimensions, see Ref.~\cite{Seiberg}.
In this work it was shown, on the basis of
strong coupling analysis, that
there is a cusp in the partition function of the $CP(N-1)$ model
at $\theta=\pi$, implying that the
expectation value of the two-dimensional
topological charge does not vanish at this point. This tells us
that $CP$-invariance is dynamically
spontaneously broken at $\theta=\pi$.

The above result is in full agreement with Witten's picture
of the vacuum family in the $CP(N-1)$ model,
with $N$ states --- one global minimum, other local ones ---
entangled in the $\theta$ evolution. At $\theta =\pi$
two minima are degenerate, but they are characterized by opposite values of
the topological charge VEV's,
$$
\langle \varepsilon_{\alpha\gamma} \partial^\alpha  
n_{\ell}^*\, \partial^\gamma n^{\ell}\rangle
= \pm
\Lambda^2\,.
$$
The kink (confined monopole) can be viewed as 
a barrier separating two domains (two degenerate strings)
carrying opposite $CP$.

On the other hand, the bulk four-dimensional theory is weakly coupled,
and for each given $\theta$ the bulk vacuum is unique.
There is no spontaneous
$CP$ violation  in the four-dimensional bulk theory at $\theta=\pi$.
One can easily check this assertion by carrying out a direct instanton calculation.

\section{Fusing strings}

As has been already mentioned, in QCD one can consider
not only basic strings, but 2-strings, 3-strings, ... , $k$-strings,
and their excitations. $k$-strings are composite flux tubes
attached to color sources with $N$-ality $k$. 
Moreover, the $N$-string ensembles --- i.e. $N$-strings ---
can decay into a no-string state. It is natural to ask how
these phenomena manifest themselves in the model under consideration.

If the {\em ansatz} (\ref{ansa}) defines a basic string,
it is not difficult to generalize this definition
to get an analog of 2-strings, 3-strings, etc., for instance, 
\beq
\Phi_{\rm 2-string} = \sqrt{\xi}\,{\rm diag} \, (e^{i\alpha (x) }, \, e^{i\alpha (x) }
\, ,1, ... ,1)\,,
\quad x\to\infty \,.
\label{ansa2}
\eeq
The solution (\ref{ansa2})
breaks SU($N$) symmetry 
down  to U(1)$\times$SU$(2)$$\times$SU$(N-2)$  (at $N>3$).
This means that the world-sheet (two-dimensional) theory of 
the   string moduli
is the SU($N$)/(U(1)$\times$SU$(2)$$\times$SU$(N-2)$) sigma model.
This is also known as the Grassmannian $G_{2,N}$ model.
At large $N$ it has more fields, by a factor of 2,
than the $CP(N-1)$ model; other features are quite similar.

The statement that in our model the world-sheet theory for $k$-strings
is the Grassmannian $G_{k,N}$ model has   a clear-cut indirect
confirmation. Indeed, the $k$-string {\em ansatz} of the type 
indicated in Eq.~(\ref{ansa2}) tells us that the number of distinct
classical strings is 
\beq
\nu (k,N)= C^N_k = \frac{N!}{ k! (N-k)! }\,,
\label{numv}
\eeq
since $k$ phase factors
$e^{i\alpha}$ can be distributed arbitrarily in $N$ positions.
{}From the  two-dimensional perspective
this number should match the number of distinct vacua
of the world-sheet theory. The latter was calculated in supersymmetric
$G_{k,N}$ model in Ref.~\cite{CecV}, where it was shown
to be $C^N_k$, as in Eq.~(\ref{numv}). In supersymmetric
$G_{k,N}$ model all these vacua are degenerate, i.e. we have degenerate
strings. Introducing supersymmetry breaking we move away from 
the degeneracy. In non-supersymmetric
$G_{k,N}$ model, the number  $\nu (k,N)=  C^N_k$ gives the 
number of states
in the vacuum family: the genuine vacuum plus metastable ones
entangled with the genuine vacuum in the $\theta$ evolution.

As soon as string  tensions in our model are classically determined by their
U(1) charges the tension of $k$-string is given by
\beq
T_k\, =\, 2\pi\,k\,\xi +O(\Lambda^2),
\label{kten}
\eeq
where corrections of order of $\Lambda^2$ are induced by the quantum effects
in the effective world sheet theory.

If we add up $N$ strings, the resulting conglomerate is connected to the 
ANO string.

\section{Kinks are confined monopoles}

The $ CP(N-1)$ models are asymptotically free theories and flow
 to strong coupling  in the infrared. Therefore, 
 the non-Abelian strings discussed in the previous
sections are in a highly quantum regime. To make contact with
the classical Abelian strings we can introduce parameters which 
explicitly break
the diagonal color-flavor SU(N)$_{{\rm diag}}$ symmetry lifting 
the orientational string moduli. This allows us to obtain a quasiclassical 
interpretation of the confined monopoles as string junctions,
and follow their evolution from (almost) 't Hooft--Polyakov
monopoles to highly quantum sigma-model
kinks. In the supersymmetric case this was done in 
Refs.~\cite{Tong,ShifmanYung, HananyTong}.

\subsection{Breaking SU($N$)$_{\rm diag}$}

In order to trace the monopole evolution
we modify our basic model
(\ref{redqed})
introducing, in addition to the already existing fields, a complex adjoint 
scalar field $a^a$,
\beqn
S &=& \int {\rm d}^4x\left\{\frac1{4g_2^2}
\left(F^{a}_{\mu\nu}\right)^{2}
+ \frac1{4g_1^2}\left(F_{\mu\nu}\right)^{2}
+\frac1{g^2_2}|D_{\mu}a^a|^2
 \right.
 \nonumber\\[3mm]
&+&
 {\rm Tr}\, (\nabla_\mu \Phi)^\dagger \,(\nabla^\mu \Phi )
+\frac{g^2_2}{2}\left[{\rm Tr}\,
\left(\Phi^\dagger T^a \Phi\right)\right]^2
 +
 \frac{g^2_1}{8}\left[ {\rm Tr}\,
\left( \Phi^\dagger \Phi \right)- N\xi \right]^2 
 \nonumber\\[3mm]
 &+&\left.
\frac12 {\rm Tr}\,\left|a^a T^a\,\Phi+\Phi\,\sqrt{2}M\right|^2
+\frac{i\,\theta}{32\,\pi^2} \, F_{\mu\nu}^a \tilde{F}^{a\,\mu\nu}
 \right\}\,,
\label{model}
\eeqn
where $D_{\mu}$ is a covariant derivative acting in the adjoint
representation of SU($N$) and $M$ is a mass matrix for scalar quarks $\Phi$.
We assume that it has a diagonal form
\beq
M=
\left(
\begin{array}{ccc}
m_1 & ... &  0\\[2mm]
...&...&...\\[2mm]
0&  ... & m_N \\
\end{array}
\right)\,,
\label{massmat}
\eeq
with the vanishing sum of the diagonal entries, 
\beq
\sum_{A=1}^N m_A =0\,.
\label{vsde}
\eeq
Later on it will be convenient to make a specific choice of the
parameters $m_A$, namely,
\beq
M= m \times {\rm diag} \left\{  e^{2\pi i/N} , \,e^{4\pi i/N},\, ... ,
 \,e^{2(N-1)\pi i/N} , \, 1\right\} \,,
\label{spch}
\eeq
where $m$ is a single common parameter, and the constraint (\ref{vsde})
is automatically satisfied. We can (and will) assume $m$ to be real and positive.

In fact, the model (\ref{model}) presents  a less reduced bosonic part 
of the \ntwo super\-symmetric
theory than the model (\ref{redqed}) on which we dwelled above. 
In the \ntwo supersymmetric
theory the adjoint field is a part of \ntwo  vector multiplet.
For the purpose of the string solution the field $a^a$
is sterile as long as $m_A=0$.
Therefore, it could be  and was ignored in the previous sections.
However, if one's intention is to connect oneself to the
quasiclassical regime, $m_A\neq 0$, and the adjoint field must be reintroduced.

For the reason which will become clear shortly, let us assume that, 
although  $m_A\neq 0$, they are all small compared to
$\sqrt{\xi}$, 
$$m  \ll \sqrt{\xi}\,,$$
but $m\gg \Lambda$.
For generic non-degenerate values of $m_A$ the adjoint field develops VEV's,
\beq
\langle a \rangle =-\sqrt{2}
\left(
\begin{array}{ccc}
m_1 & ... &  0\\[2mm]
...&...&...\\[2mm]
0&  ... & m_N \\
\end{array}
\right)\,.
\label{avev}
\eeq
The vacuum expectation values of the scalar quarks $\Phi$ 
remain intact; they are  given by Eq.~(\ref{diagphi}). For the particular 
choice specified in Eq.~(\ref{spch})
\beq
\langle a \rangle =-\sqrt{2}\,  m\,\, {\rm diag} \left\{  e^{2\pi i/N} , 
\,e^{4\pi i/N},\, ... , \,e^{2(N-1)\pi i/N} , \, 1 \right\} \,.
\label{aspch}
\eeq

Clearly the diagonal
color-flavor group SU($N$)$_{\rm diag}$ is now broken by adjoint VEV's
down to U(1)$^{N-1}\times Z_N$.
Still, the solutions for the Abelian (or $Z_N$) strings are the same 
as was discussed in Sect.~\ref{search} since the adjoint
field does not enter these solutions. In particular,  we  have $N$
distinct $Z_N$   string solutions depending on what particular squark winds
at infinity, see Sect.~\ref{search}. Say, the string solution with the 
winding last flavor  is 
still given by Eq.~(\ref{znstr}).

What is changed with the color-flavor   SU($N$)$_{\rm diag}$  
explicitly broken by $m_A\neq 0$,
the rotations (\ref{nastr}) no more generate zero modes. 
In other words, the fields
$n^{\ell}$ become quasi-moduli:
a shallow potential for the quasi-moduli $n^l$ on the string world sheet
is generated. 
This potential is shallow as long as $m_A\ll \sqrt\xi$.
 
This potential
was calculated in the $CP(1)$ case in Ref.~\cite{ShifmanYung}; the 
$ CP(N-1)$
case was treated in \cite{HananyTong}. It has the following form:
\beq
V_{CP(N-1)}=2\beta\left\{\sum_l |m_l|^2|n^l|^2-\left|\sum_l m_l|n^l|^2
\right|^2\right\}\,.
\label{pot}
\eeq
The potential simplifies if the mass terms are chosen according to
(\ref{spch}),
\beq
V_{CP(N-1)}=2\beta\, m^2 \left\{
1
-\left|
\sum_{\ell=1}^N \, e^{2\pi i\, \ell/N }\,  | n^\ell |^2
\right|^2\right\}\,.
\label{pots}
\eeq
This potential is obviously invariant under the cyclic
substitution
\beq
\ell\to\ell +k\,,\qquad    n^\ell \to n^{\ell + k}\,,\qquad \forall\,\,\ell\,,
\label{cycle}
\eeq
with $k$ fixed. This property will be exploited below.

Now our effective two-dimensional theory on the string world sheet becomes
a massive $CP(N-1)$ model. The potential (\ref{pot}) or (\ref{pots}) has $N$ vacua
at 
\beq
n^\ell=\delta^{\ell\ell_0}\,, \qquad \ell_0 =1, 2, ... , N\,.
\label{nell}
\eeq
These vacua correspond
to $N$ distinct Abelian $Z_N$ strings with $\vp^{\ell_0 \ell_0}$ winding at 
infinity, see Eq.~(\ref{str}).

\subsection{Evolution of monopoles}
\label{eom}

\begin{figure}
\epsfxsize=9cm
\centerline{\epsfbox{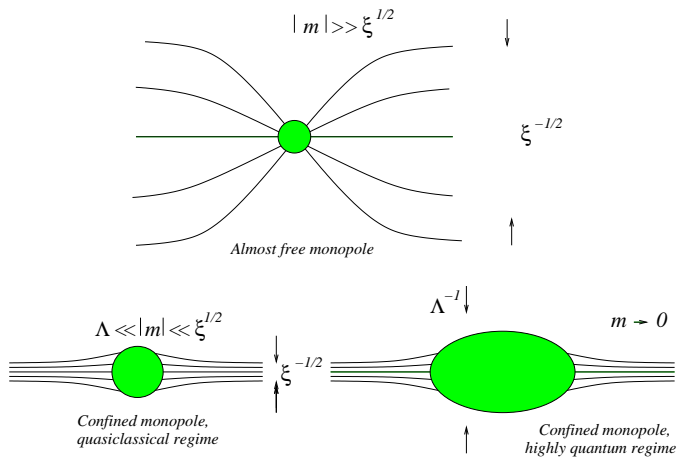}}
\caption{
Evolution of the confined monopoles.}
\label{sixf}
\end{figure}

Our task in this section is to trace the evolution of the confined monopoles
starting from the quasiclassical regime, deep into the quantum regime.
For illustrative purposes it will be even more instructive if we start from the
limit of weakly confined monopoles, when in fact they present
just slightly distorted 't Hooft-Polyakov monopoles (Fig.~\ref{sixf}).
For simplicity, in this section we will set $\theta =0$.
To further simplify the subsequent
discussion  we will 
not treat $N$ as a large parameter in this section, i.e. we will 
make no parametric distinction between $m$ and $mN$.

Let us start from the limit $| m_A| \gg \sqrt{\xi }$ and take all
masses of the same order,  as in Eq.~(\ref{spch}). 
In this limit the scalar quark expectation values can be neglected,
and the vacuum structure is determined by VEV's of the adjoint
$a^a$ field. In the non-degenerate case
the gauge symmetry SU($N$) of our microscopic model
is broken down to U(1)$^{N-1}$ modulo possible discrete subgroups.
This is the text-book situation for occurrence of the
 SU($N$) 't Hooft-Polyakov monopoles. The monopole core size
 is of the order of $| m|^{-1}$. 
The 't Hooft-Polyakov
solution remains valid up to much larger distances of the order
of $\xi^{-1/2}$. At distances larger than $\sim \xi^{-1/2}$  
the quark VEV's become important. As usual, the U(1) 
charge condensation leads
to the formation of the U(1) magnetic flux tubes, with the transverse size
of the order of $\xi^{-1/2}$ (see the upper picture in Fig.~\ref{sixf}).
The flux is quantized; the flux tube tension is tiny in
the  scale of  the square
of the monopole mass. Therefore, what we deal in this limit is
basically a very weakly confined 't Hooft-Polyakov monopole.

Let us verify that the confined monopole is a junction of two strings.
Consider the junction of two $Z_N$ strings   corresponding to two 
``neighboring'' vacua of the $CP(N-1)$ model. For $\ell_0$-th vacuum $n^{\ell}$ 
is given by (\ref{nell}) while for $\ell_0 +1$-th vacuum it is given by the 
same equations with $\ell_0\to\ell_0 +1$. The flux of this junction is 
given by the difference of the fluxes of these two strings. Using (\ref{str})
we get that the flux of the junction is
\beq
  4\pi\,\times \, {\rm diag} \, \frac12\, \left\{  ...\, 0, \,1 ,\,  -1,\, 0
,\, ... \right\} \,
\label{monflux}
\eeq
with the non-vanishing entries located at 
positions $\ell_0$ and $\ell_0 +1$.
These are  exactly the  fluxes of $N-1$ distinct 't Hooft-Polyakov 
monopoles occurring in the 
SU($N$) gauge theory provided that SU($N$)
is spontaneously broken down to U(1)$^{N-1}$. We see that 
in the quasiclassical limit of large $| m_A|$ the Abelian 
monopoles play the role of junctions
of the Abelian $Z_N$ strings. Note that in various models 
the fluxes of monopoles and 
strings were shown \cite{Bais,PrVi,MY,Kn,ABEK} to match each other
so that the monopoles can be confined by strings in the Higgs phase.
The explicit solution for the confined monopole as a 1/4 BPS junction
of two strings was obtained in \cite{ShifmanYung} for $N=2$ case
in \ntwo supersymmetric theory. The general solution for 1/4 BPS junctions
of semilocal strings was obtained in \cite{INOS}.

Now, if we   reduce $\left|  m\right| $,
$$
\Lambda \ll \left|  m\right|   \ll \sqrt{\xi}\, ,
$$
the size of the monopole ($\sim \left|  m\right|^{-1} $) becomes
larger than the transverse size of the attached strings.
The monopole gets squeezed  in earnest by
the strings --- it becomes  a {\em bona fide} confined
monopole (the  lower left corner of  Fig.~\ref{sixf}).
A macroscopic description of such monopoles is provided
by  the massive $CP^{N-1}$ model, see Eq. (\ref{pot}) or (\ref{pots}).
The confined monopole is nothing but
the massive sigma-model kink.

As we further diminish $\left|  m\right|$
approaching $\Lambda$ and then getting  below $\Lambda$,
the size of the monopole grows, and, classically, it would explode.
This is where quantum effects in the world-sheet theory take over.
This domain presents the regime of
highly quantum world-sheet dynamics.
While the thickness of the string (in the transverse direction) is
$\sim \xi ^{-1/2}$, the
$z$-direction size of the kink  representing the confined
monopole in the highly quantum regime is much larger, 
$\sim \Lambda^{-1}$,
see the  lower right corner in  Fig.~\ref{sixf}.
In passing from $ m \gg \Lambda$ to 
$ m \ll \Lambda$ we, in fact, cross a line of the phase transition
from Abelian to non-Abelian strings.   
This is discussed in Sect.~\ref{atna}.

\section{Abelian to non-Abelian string phase transition}
\label{atna}

In this section we will restrict ourselves to the choice of the mass parameters
presented in Eq.~(\ref{spch}). Correspondingly, the
potential of the massive $CP^{N-1}$ model 
describing the  quasimoduli has the form (\ref{pots}).

At large $m$, $m\gg\Lambda$,  the model is at weak coupling, so the 
quasiclassical analysis is applicable.
$N$ quasiclassical vacua are presented in Eq.~(\ref{nell}).
The invariance of $V_{CP(N-1)}$ under the cyclic permutations
(\ref{cycle}) implies a $Z_N$ symmetry of the world-sheet theory of the quasimoduli. 
In each given vacuum
the $Z_N$ symmetry is spontaneously
broken. $N$ vacua have strictly degenerate vacuum energies, which, as 
we already know, leads to the 
kinks   {\em de}confinement. From the four-dimensional point of view this
means that we have $N$ strictly degenerate Abelian strings (the $Z_N$ strings).

The flux of the Abelian 't Hooft-Polyakov monopole 
equals to the difference of the fluxes of two ``neighboring'' strings,
see (\ref{monflux}).  Therefore, the confined monopole
in this regime is obviously  a junction of two distinct
$Z_N$-strings. It is seen as a quasiclassical kink
interpolating between the ``neighboring'' $\ell_0$-th and $(\ell_0 +1)$-th
vacua  of the 
effective massive $CP(N-1)$ model on the string world sheet. 
A  monopole can move freely along the string as both 
attached strings are tension-degenerate.

Now if we further reduce $m$ tending  it to zero, the picture changes. 
At $m=0$ the global symmetry
SU($N$)$_{\rm diag}$ is unbroken, and so is the 
discrete $Z_N$ of the massive 
$CP(N-1)$ model with the potential (\ref{pots}).
$N$ degenerate vacua of the quasiclassical regime 
give place to $N$ non-degenerate ``vacua" depicted in Fig.~\ref{odin}
(see Sect.~\ref{Dyn}). The fact that $\langle n^\ell\rangle =0$
in the quantum regime signifies that in the limit $m\to 0$
the $Z_N$ symmetry of the massive model gets restored.
Now kinks are confined, as we know from  Sect.~\ref{Dyn}. 

{}From the standpoint of the four-dimensional 
microscopic theory the tensions of $N$ 
non-Abelian strings get a split, and the non-Abelian monopoles, 
in addition to  the four-dimensional confinement 
(which ensures that the monopoles
are attached to the strings) acquire a two-dimensional confinement along 
the string:
a monopole--anti-monopole   forms a meson-like configuration, with necessity, 
see Fig.~\ref{tri}.

\begin{figure}
\epsfxsize=11cm
\centerline{\epsfbox{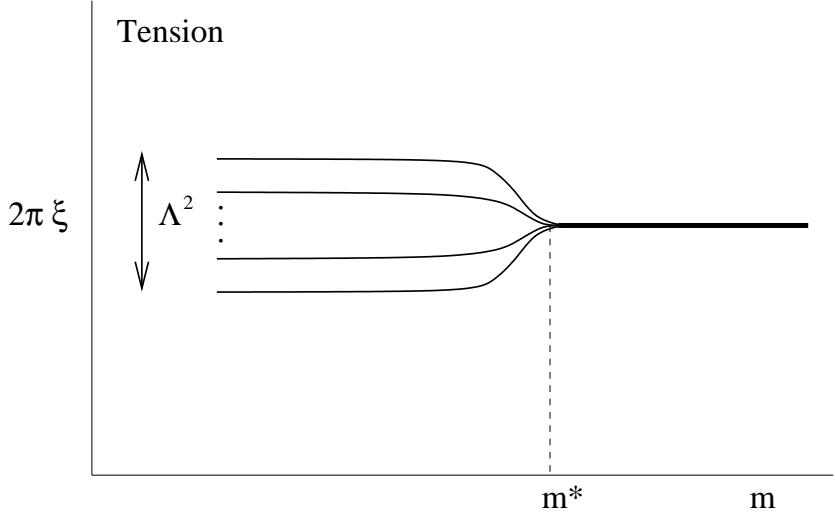}}
\caption{Schematic dependence of string tensions on the mass parameter $m$.
At small $m$ in the non-Abelian confinement phase the tensions are split
while in the Abelian confinement phase at large $m$ they are degenerative.}
\label{phtrans}
\end{figure}

Clearly these two regimes at large and  small $m$ are separated by the 
phase transition at some critical value $m_*$.
We interpret this as a phase transition
between the Abelian and non-Abelian confinement. In the Abelian confinement phase 
at large $m$, the $Z_N$ symmetry is spontaneously broken, all $N$ strings 
are strictly degenerate, and there is no two-dimensional confinement of
the 4D-confined monopoles. Instead,
in the non-Abelian confinement phase occurring at small $m$, the
$Z_N$ symmetry is fully restored, all $N$ elementary strings are split, and 
the 4D-confined monopoles combine with anti-monopoles to form a 
meson-like configuration
on the string, see Fig.~\ref{tri}. We show schematically
the dependence of the string tensions on $m$ in these two phases in 
Fig.~\ref{phtrans}. 

It is well known \cite{W93}
that two-dimensional $CP(N-1)$ model can be obtained
as a low-energy limit of a U(1) gauge theory with  $N$ flavors of 
complex scalars $n^{\ell}$ and the potential
\beq
e^2\beta^2 \left(|n^{\ell}|^2-1\right)^2,
\label{gsigmod}
\eeq
where $e^2$ is U(1) gauge coupling. Classically 
the $CP(N-1)$ model  corresponds to
the Higgs phase of this gauge theory. The potential (\ref{gsigmod})
forces $n^{\ell}$ to develop VEV's breaking the U(1) gauge 
symmetry. Then the U(1) photon
becomes heavy and can be integrated out. Namely, in the low-energy limit
the gauge kinetic term can be ignored which leads us to the model (\ref{one}).

To include the masses $m_A$ in this theory we 
add, following \cite{W93}, a neutral complex scalar field $\sigma$ and 
consider the U(1) gauge theory
with the potential
\beqn
S^{(1+1)}
&=&
\int d t\, dz \,  \left\{2 \beta\,|\nabla_{\alpha}\, n |^2
+\frac{1}{4e^2}F_{\alpha\gamma}^2 + \frac1{e^2}|\pt_\alpha \sigma|^2
\right.\,,
\nonumber\\[3mm]
&+& \left.
4\beta\,\left|\left(\sigma-\frac{m_{\ell}}{\sqrt{2}}\right)
n^{\ell}\right|^2+
2e^2\beta^2 \left(|n^{\ell}|^2-1\right)^2\right\},
\label{2dgauge}
\eeqn
where $\nabla_\alpha=\pt_\alpha-iA_\alpha$ ($A_\alpha$ is the 
two-dimensional U(1) gauge potential).

At large $m_A$ this theory is in the Higgs phase. Moreover, quantum effects
do not destroy the Higgs phase because 
the coupling constant is small.  Namely,
$\sigma$ develops a VEV,
$$
\langle \sigma\rangle =
m_{\ell_0}\,,
$$ 
while VEV's of $n^{\ell}$ 
are given by (\ref{nell}). In this phase both the U(1) gauge field and 
the scalar field
$\sigma$ become heavy and can be integrated out leading to the 
massive $CP(N-1)$ model with the potential (\ref{pot}).

At small $m_A$  this theory is in the Coulomb phase. The VEV's of
$n^{\ell}$ vanish, and the photon becomes massless. 
Since the Coulomb potential
in two dimensions is linear, the photon masslessness
results in   confinement of kinks \cite{5}.
Thus, the phase transition which we identified above, separates
the Higgs and Coulomb phases of the  two-dimensional U(1) gauge theory
(\ref{2dgauge}). The Higgs phase is characterized by a
broken $Z_N$ symmetry and degenerate vacua, while in  the
Coulomb phase the $Z_N$ symmetry
gets restored,  and the  vacua split.  In four dimensions the
former phase is an Abelian confinement phase with degenerate
Abelian strings and 2D deconfinement of monopoles. The latter phase
is a non-Abelian confinement phase with $N$ split non-Abelian strings
and  non-Abelian 2D-confined monopoles forming meson-like
configurations on these strings. Note that the description of
the $CP(N-1)$
theory on the string world sheet as a U(1) gauge theory (\ref{2dgauge})
was used in \cite{HananyTong} in a supersymmetric setting.

In particular,
we expect that in the $N=2$ case
the massive $CP(1)$ model is in the same universality class
as the two-dimensional Ising model. Therefore, we conjecture that the
phase transition from the Abelian  confinement phase to the non-Abelian 
one is of the second order, and is described
(at $N=2$) by   conformal field theory
with the central charge $c=1/2$, which corresponds to a
free Majorana fermion.

To conclude this section we would like to stress that
we encounter a crucial difference between the non-Abelian confinement
in supersymmetric and non-supersymmet\-ric gauge theories.
For BPS strings in
supersymmetric theories we do not have a phase transition separating
the phase of the non-Abelian stings from that of the Abelian strings
\cite{ShifmanYung,HananyTong}. Even for small values of
the mass parameters supersymmetric theory
strings are strictly degenerate, and the $Z_N$ symmetry is
spontaneously broken. In particular, at $m_A=0$ the order parameter
for the broken $Z_N$, which
differentiates  $N$ degenerate vacua of the supersymmetric $CP(N-1)$
model, is the bifermion condensate of two-dimensional fermions living
on the string world sheet of the non-Abelian BPS string. 

An example of the  deconfinement phase transition at a critical mass is known
\cite{gvy} in four-dimensional softly broken ${\cal N}=2$  
SQCD; in this model the order parameter
is the Seiberg-Witten monopole condensate, and the collision of vacua
happens in the parameter space, which is absent in our model.
Note, that in some two-dimensional
supersymmetric  theories both  Coulomb and Higgs 
branches are present and they
have distinct $R$ symmetries and  different renormalization group
flows in the infrared domain \cite{W97}. Interpolation between two branches
is a rather delicate issue since the transition region is described by a
nontrivial geometry in the moduli space. A recent analysis of the 
supersymmetric case \cite{Ooguri} shows that the
two phases can even coexist on the world sheet and, moreover,
integration over the form of the boundary is necessary to make the theory
self-consistent. 

We do {\em not} expect such a picture in 
the non-supersymmetric case under consideration.

\section{The SU(2)$\times$U(1) case}
\label{su2c}

The $N=2$ case is of special importance, since the
corresponding world-sheet theory, $CP(1)$,
is exactly solvable. 
In this section we discuss   special features of this theory
in more detail.  The Lagrangian  on the string world sheet is 
\beqn
S^{(1+1)}
&= &
2 \beta\,  \int d t\, dz \,\left\{ 
(\pt_\alpha\, n^*\pt_\alpha\, n) + (n^*\pt_\alpha\, n)^2
+m^2\left[1-\left(|n^1|^2-|n^2|^2\right)^2\right]
\right\}
\nonumber\\[4mm]
&-& \frac{\theta}{2\pi}\,\int d t\, dz \,\varepsilon_{\alpha\gamma}\left(
\pt_\alpha\, n^*\, \pt_\gamma\, n \right)
\,,
\label{cp1}
\eeqn
where in the case at hand the mass parameter $m=m_1=-m_2$, see (\ref{spch}).
In this theory the mass term breaks SU(2)$_{\rm diag}$ symmetry down 
to U(1)$\times$ Z$_2$ 
since the potential is invariant under the exchange
$n^1\leftrightarrow n^2$. It has two minima: the first one  located 
at $n^1=1$, $n^2=0$, and the second minimum at $n^1=0$, $n^2=1$.

Now let us discuss the $m=0$ limit, i.e. non-Abelian strings,
in more detail.  Setting $N=2$ we arrive at a non-Abelian string
with moduli forming a
$CP(1)$ model on the world sheet. The very same 
string emerges in
the supersymmetric model \cite{Markov}
which supports non-BPS string solutions. 
It is instructive to discuss how the pattern we have established for the
$CP(N-1)$ string is implemented in this case.

\begin{figure}
\epsfxsize=11cm
\centerline{\epsfbox{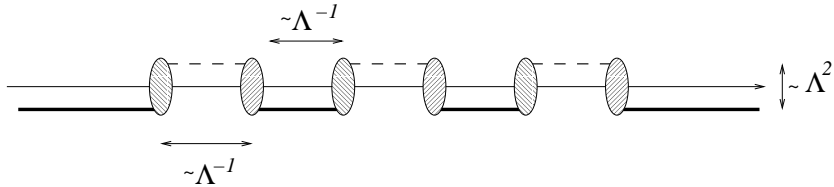}}
\caption{
Breaking of a would-be string through the kink-anti-kink pair creation in
$CP(1)$.  Thick solid line shows the energy density of the true vacuum,
while dashed one indicates the energy density of the ``metastable"
vacuum.}
\label{figSBone}
\end{figure}

Unlike $CP(N-1)$, the
$CP(1)$ model has only one parameter, the dynamical scale
$\Lambda$. The small expansion parameter $1/N$ is gone.
Correspondingly, the kink-anti-kink  interaction becomes strong,
which invalidates quasiclassical-type  analyses. On the other hand,
this model was exactly solved \cite{SBone}.
The exact solution shows that the SU(2) doublets
(i.e. kinks and anti-kinks) do not show up in the physical spectrum,
and the only asymptotic states present in the  spectrum are
SU(2) triplets, i.e.  bound states of kinks and anti-kinks.
(Note that there are no bound states of the SU(2)-singlet type).
As was noted by Witten \cite{5} passing from large $N$
to $N=2$ does not change the picture qualitatively.
In the quantitative sense it makes little sense now to speak
of the kink linear confinement, since there is no suppression
of the string breaking. The metastable vacuum 
entangled with the true vacuum in the $\theta$ evolution,
is, in fact, grossly unstable. Attempting to create a long string,
one just creates multiple kink-anti-kink pairs, as shown in Fig.
\ref{figSBone}. We end up with pieces of broken string
of a typical length $\sim \Lambda^{-1}$.

There is a special interval of $\theta$ where long strings
do exist, however, and, hence, we can apply the approach developed in
the previous sections to obtain additional information.
The $CP(1)$ model at $\theta=\pi$   turns out to be integrable 
\cite{Zam,Fat},
much in the same way as at $\theta =0$. From the exact 
solution \cite{Zam,Fat} it is known that
at $\theta =\pi$ there 
are no localized asymptotic states in the physical spectrum ---
the model becomes conformal.
The exact solution confirms the presence of deconfined kinks
(doublets) at $\theta=\pi$
and their masslessness. The $S$-matrix for the scattering
of these massless states has been found in \cite{Zam,Fat}.

We will focus on a small interval
of $\theta$ in the vicinity of $\theta =\pi$. It is convenient to
introduce a new small parameter
\beq
\varepsilon = |\pi - \theta |\,.
\label{ma}
\eeq 
If $\varepsilon \ll 1$, our model again becomes
two-parametric. We will argue that in this regime
the string tension in  the $CP(1)$ model is
\beq
\Delta T_{CP(1)} \sim \Lambda^2\,  \varepsilon\,,
\label{maa}
\eeq
while the kink mass and the string size scale as
\beq
M_n \sim \Lambda \,  \varepsilon^{1/2}\,,\qquad L \sim \Lambda^{-1}
\,  \varepsilon^{-1/2}\,.
\label{maaa}
\eeq
The mass of the kink-anti-kink bound state
also scales as
\beq
M \sim \Lambda \,  \varepsilon^{1/2}\,,
\label{maaam}
\eeq
so that at $\theta =\pi$ the string tension vanishes
allowing the model to become conformal.

Let us elucidate the above statements starting from the string tension.
In the $CP(1)$ model the vacuum family
consists of two states: one true vacuum, and another
--- local --- minimum,  a companion of the true vacuum
in the  $\theta$ evolution. This fact can be confirmed
by consideration of the supersymmetric
$CP(1)$ model which has two degenerate vacua.
Upon a soft SUSY breaking deformation, a small fermion mass
term, the above vacua split: one minimum moves to a higher energy
while another to a lower one. The roles of these 
non-degenerate minima interchange in the process of the $\theta$
evolution from zero to $2 \pi$;
at $\theta = \pi$ they get degenerate.

Returning to the the non-supersymmetric $CP(1)$ model,
it is not difficult to derive that in the vicinity of
$\theta = \pi$ the vacuum energy densities ${\cal E}_{1,2}$
of the two vacua behave as
\beq
{\cal E}_{1,2} = {\cal E}_{0} \pm \Lambda^2 \left(\theta -\pi\right)\,.
\label{maaamm}
\eeq
This formula proves that the difference of
the vacuum energy densities (a.k.a. the string tension)
scales as indicated in Eq.~(\ref{maa}).

Now the validity of Eq.~(\ref{maaa}) can be checked
with ease. Indeed, the kink momentum (which is $\sim L^{-1}$)
is of the order of its mass. Therefore, the kink and the anti-kink
in the bound pair are right at the border
of non-relativistic and ultrarelativistic regimes.
No matter which formula for their potential energy we use, we get
$E_n \sim \Lambda \varepsilon^{1/2}$, so that the potential energy 
of the bound state is of the order of the kinetic 
energy of its constituents. The total mass of the bound state is
then given by Eq.~(\ref{maaam}).
This is in full agreement with the fact \cite{Fat} that
the conformal theory one arrives at in the limit
$\theta =\pi$ has the Virasoro central charge $c=1$.
At $c=1$ the spectrum of the scaling dimensions
is given by $(1/4)\times$(integer)$^2$.

It is easy to verify that any regime other than (\ref{maaa})
is inconsistent. Here we note in passing that
our result contradicts the analyses of Refs.~\cite{Affl,Cont}.
In these papers a deformation of the exact $\theta =\pi$
solution of the $CP(1)$ model was considered, with the conclusion
that $M\sim \Lambda\,\varepsilon^{2/3}$. This scaling regime 
is in contradiction with our analysis.

An alternative analysis of the $CP(1)$ model
at generic $\theta$ can be carried out 
using the quasiclassical picture developed by Coleman a
long time ago \cite{Coleman}. Namely, in the dual fermionic version
of the model $\theta$ corresponds to the constant electric field
created by two effective charges located at the ends of the strings.
The value of the electric field experiences  jumps at the  kinks'
positions, since the kinks are charged too. Generically the system is in
2D Coulomb phase, with the vanishing photon mass.  Coleman's 
analysis  is qualitatively 
consistent with the description of the $CP(N-1)$ model as a Coulomb 
phase of the U(1) gauge theory (\ref{2dgauge}) reviewed in Sect.~\ref{atna}
and with the solution  \cite{5} of the $CP(N-1)$ models at large $N$.

\section{Dual picture}
\label{dupi}

It is instructive to compare properties of the QCD strings 
summarized at the end
of Sect.~\ref{introd} with those emerging in the model under 
consideration.
First of all, let us mention the most drastic distinction.
In QCD, the string tension, excitation energies, and all other
dimensionful parameters are proportional to the only scale of the 
theory,
the dynamical scale parameter $\Lambda_{\rm QCD}$. In the model at hand we have two
mass scales, $\sqrt\xi$ and $\Lambda$. To ensure full theoretical 
control we must assume that $\xi \gg \Lambda^2$.

The transverse size of the string under consideration
is proportional to $1/\sqrt\xi$. Correspondingly, a large component
in the string tension is proportional to $\xi$, see Eq.~(\ref{splitten}). 
It is only a fine structure
of the string that is directly related to $\Lambda$,
for instance, the splittings between the excited strings and
the string ground state. The decay rates of the excited strings
are exponentially suppressed,   $\sim \exp(-\gamma N^2)$ 
in the QCD case and $\sim \exp(-\gamma N )$  
in our model. The confined monopoles in our model
are in one-to-one correspondence with gluelumps of QCD
(remember, in the model at hand we deal with the Meissner effect, while
it is the {\em dual} Meissner effect that is operative in QCD).

$N$ strings in QCD can combine to produce a no-string state,
while $N$ non-Abelian strings in our model can combine to produce an 
ANO string,
with no structure at the scale $\Lambda$. The only scale of the ANO
string is $\xi$.

Moreover, confinement in our model should be thought of
as dual to  confinement in pure Yang--Mills theory with {\em no sources}
because there are no
monopoles attached to  the string ends  in our model. If we started 
from a
SU($N+1$) gauge theory spontaneously broken to
SU($N$)$\times$U(1) at a very high scale,
then in that theory there would be extra very heavy monopoles
that could be attached to the ends of
our strings. However, in the SU($N$)$\times$U(1) model {\em per se}
these very heavy monopoles  become
infinitely heavy. The SU($N$) monopoles we have considered in the 
previous
sections  are junctions of  two elementary
strings   dual to gluelumps, rather than to the end-point sources.

Our model exhibits a phase transition in $m$ between  the Abelian and
non-Abelian types of confinement. As well-known \cite{DS,matt},
the Abelian confinement
leads to proliferation of hadronic states: the bound state 
multiplicities
within the Abelian confinement are much higher than they ought to
be in QCD-like theories.

In our model the Abelian confinement regime occurs at large $m$,
($m\gg\Lambda$). In this region we have $N$
degenerate $Z_N$  strings with the tensions given in (\ref{ten}).
If we extend our model to introduce superheavy monopoles  (see above)
as the end-point source objects, it is the $N$-fold degeneracy
of the $Z_N$ strings occurring in this phase that
is responsible for an excessive multiplicity of the ``meson"
states.

Now, as we  reduce $m$ and eventually cross  the phase transition point
$m_*$, so that $m <m_*$,
the strings under consideration become non-Abelian. The
world-sheet $CP(N-1)$ model becomes  strongly coupled,
and  the string tensions
split according to Eq. (\ref{splitten}). The splitting
is determined by the $CP(N-1)$ model and is $\sim \Lambda^2/N$. 
Thus,
(at $\theta=0$) we have  one lightest string --- the ground state --- 
as expected in QCD.
Other $N-1$ exited strings become metastable. They are
connected to the ground-state string through the monopole-anti-monopole
pairs.
At large $N$ their
decay rates are  $\sim \exp{(-N)}$.

\begin{figure}
\epsfxsize=3cm
\centerline{\epsfbox{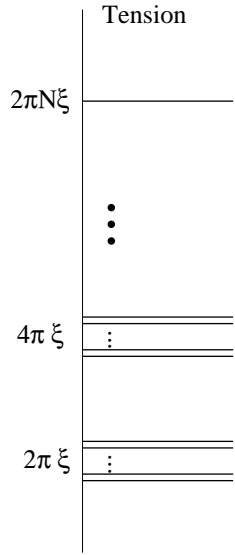}}
\caption{
The string spectrum in the non-Abelian confinement phase. The $k$-strings
at each level are split.}
\label{spectrum}
\end{figure}

Besides  $N$ elementary strings we also have $k$-strings which can be
considered as a bound states of $k$ 
elementary strings. Their tensions are given
in Eq.~(\ref{kten}). At each level $k$ we have $N!/k!\,(N-k)!$
split strings. The number of strings at the level
$k$ and $N-k$ are the same. At the highest  level $k=N$
we have only the ANO string.  The string spectrum in our theory is shown 
in Fig.~\ref{spectrum}.

The dual of this phenomenon is the occurrence of the
$k$-strings in QCD-like theories.
If $k\gg 1$ we have a large number of metastable strings, with 
splittings suppressed by inverse powers of $N$, which are connected to the
ground state string through a gluelump.

At small $N$ all metastable strings become 
unstable and  practically unobservable.

\section{Conclusions}

Our main task in this work was developing a simple reference
set-up which supports non-Abelian strings and confined monopoles 
at weak coupling. We construct a simple non-supersymmetric SU$(N)\times$U(1)
Yang--Mills theory which does the job. The advantages of the large-$N$
limit (i.e. $1/N$ expansion) are heavily exploited.
We discover, {\em en route},
a phase transition between Abelian and non-Abelian confinement regimes.
We discuss in detail a dual picture where the confined monopoles
turn into string-attached gluelumps; these gluelumps separate
excited strings from the ground state. The non-Abelian strings we
obtain in non-Abelian regime have many common features with QCD $k$-strings;
however, they have significant distinctions as well. At the present level 
of understanding, this is as good as it gets on the road to
quantitative theory of QCD strings.

\section*{Acknowledgments}

We are very grateful to Adam Ritz for stimulating discussions and communications. We would like to thank G. Mussardo for pointing out
an inaccuracy in referencing.

This work was carried out, in part, within the framework of
the Workshop {\sl QCD and String Theory} at  
Kavli Institute for Theoretical Physics,
UCSB, Santa Barbara, CA, August 2 -- December 17, 2004.
A.G. and M.S. are grateful to colleagues and staff of KITP 
for kind hospitality, and acknowledge financial support of the National 
Science Foundation under Grant No. PHY99-07949.
The work  of M.S. was supported in part by DOE grant DE-FG02-94ER408.
The work of A.~Y. was partially supported by the RFBF grant 
No.~05-02-17360
and by Theoretical Physics Institute at the University of Minnesota.
The work of A.~G. was partially supported by the RFBF grant No.~04-01-00646
and by Theoretical Physics Institute
at the University of Minnesota. He also thanks IHES where
a part of the work was carried out, for  kind hospitality and support.


\vspace{3mm}

\begin{thebibliography} {99}

\bibitem{thooft}
G.~'t Hooft,
%``Topology Of The Gauge Condition And 
%New Confinement Phases In Non-Abelian
%Gauge Theories,''
Nucl.\ Phys.\ B {\bf 190}, 455 (1981).
%%CITATION = NUPHA,B190,455;%%

\bibitem{mandelstam}
S.~Mandelstam,
%``Vortices And Quark Confinement In Non-Abelian Gauge Theories,''
Phys.\ Rept.\  {\bf 23}, 245 (1976).
%%CITATION = PRPLC,23,245;%%

\bibitem{sw}
N.~Seiberg and E.~Witten,
%``Electric - magnetic duality, monopole 
%condensation, and confinement in N=2
%supersymmetric Yang--Mills theory,''
Nucl.\ Phys.\ B {\bf 426}, 19 (1994),
(E) \ B {\bf 430}, 485 (1994)
[hep-th/9407087];
%%CITATION = HEP-TH 9407087;%%
%``Monopoles, duality and chiral symmetry breaking in N=2 supersymmetric QCD,''
Nucl.\ Phys.\ B {\bf 431}, 484 (1994)
[hep-th/9408099].
%%CITATION = HEP-TH 9408099;%%

\bibitem{ANO}
A.~Abrikosov, Sov.~Phys. JETP {\bf32} 1442  (1957)
[Reprinted in {\em Solitons and Particles}, Eds. C. Rebbi and G. Soliani
(World Scientific, Singapore, 1984), p. 356];\\
H.~Nielsen and P.~Olesen, Nucl.~Phys. {\bf B61} 45 (1973)
[Reprinted in {\em Solitons and Particles}, Eds. C. Rebbi and G. Soliani
(World Scientific, Singapore, 1984), p. 365].

\bibitem{DS}
M.~R.~Douglas and S.~H.~Shenker,
%``Dynamics of SU(N) supersymmetric gauge theory,''
Nucl.\ Phys.\ B {\bf 447}, 271 (1995)
[hep-th/9503163].
%%CITATION = HEP-TH 9503163;%%

\bibitem{matt}
A.~Hanany, M.~J.~Strassler and A.~Zaffaroni,
%``Confinement and strings in M{QCD},''
Nucl.\ Phys.\ B {\bf 513}, 87 (1998)
[hep-th/9707244].
%%CITATION = HEP-TH 9707244;%%

\bibitem{limited}
M.~Hindmarsh and T.~W.~B.~Kibble,
%``Beads On Strings,''
Phys.\ Rev.\ Lett.\  {\bf 55}, 2398 (1985);
%%CITATION = PRLTA,55,2398;%%
\\
A.~E.~Everett and M.~Aryal,
%``Comment On 'Monopoles On Strings.',''
Phys.\ Rev.\ Lett.\  {\bf 57}, 646 (1986);
%%CITATION = PRLTA,57,646;%%
\\
E.~Witten,
%"Superconducting strings",
Nucl. Phys. {\bf B249}, 557 (1985);
\\
M.~Hindmarsh,
%``Superconducting Cosmic Strings In Grand Unified Models,''
Phys.\ Lett.\ B {\bf 225}, 127 (1989);
%%CITATION = PHLTA,B225,127;%%
\\
M.~G.~Alford, K.~Benson, S.~R.~Coleman, J.~March-Russell and F.~Wilczek,
%``Zero Modes Of Nonabelian Vortices,''
Nucl.\ Phys.\ B {\bf 349}, 414 (1991);
%%CITATION = NUPHA,B349,414;%%
\\
M.~A.~C.~Kneipp,
%``Monopole confinement in non-Abelian theories,''
Int.\ J.\ Mod.\ Phys.\ A {\bf 18}, 2085 (2003);
%%CITATION = IMPAE,A18,2085;%%
%``Z(k) string fluxes and monopole confinement in non-Abelian theories,''
Phys.\ Rev.\ D {\bf 68}, 045009 (2003)
[hep-th/0211049];
%%CITATION = HEP-TH 0211049;%%
Phys.\ Rev.\ D {\bf 69}, 045007 (2004)
[hep-th/0308086].
%%CITATION = HEP-TH 0308086;%%

\bibitem{recent}
M.~Shifman and A.~Yung,
%{\em Domain walls and flux tubes in N = 2 SQCD: D-brane prototypes,}
Phys.\ Rev.\ D {\bf 67}, 125007 (2003)
[hep-th/0212293];
%%CITATION = HEP-TH 0212293;%%
%``Localization of non-Abelian gauge 
%fields on domain walls at weak coupling
%(D-brane prototypes II),''
Phys.\ Rev.\ D {\bf 70}, 025013 (2004)
[hep-th/0312257].
%%CITATION = HEP-TH 0312257;%%

\bibitem{Hanany}
A.~Hanany and D.~Tong,
%``Vortices, instantons and branes,''
JHEP {\bf 0307}, 037 (2003)
[hep-th/0306150].
%%CITATION = HEP-TH 0306150;%%

\bibitem{Auzzi}
R.~Auzzi, S.~Bolognesi, J.~Evslin, K.~Konishi and A.~Yung,
%{\em Non-Abelian superconductors: Vortices and confinement in N = 2
%SQCD,}
Nucl.\ Phys.\ B {\bf 673}, 187 (2003)
[hep-th/0307287].
%%CITATION = HEP-TH 0307287;%%

\bibitem{ShifmanYung}
M.~Shifman and A.~Yung,
%``Non-Abelian string junctions as confined monopoles,''
Phys.\ Rev.\ D {\bf 70}, 045004 (2004)
[hep-th/0403149].
%%CITATION = HEP-TH 0403149;%%

\bibitem{Tong}
D.~Tong,
%``Monopoles in the Higgs phase,''
Phys.\ Rev.\ D {\bf 69}, 065003 (2004)
[hep-th/0307302].
%%CITATION = HEP-TH 0307302;%%

\bibitem{HananyTong}
A.~Hanany and D.~Tong,
%``Vortex strings and four-dimensional gauge dynamics,''
JHEP {\bf 0404}, 066 (2004)
[hep-th/0403158].
%%CITATION = HEP-TH 0403158;%%

\bibitem{Markov}
V.~Markov, A.~Marshakov and A.~Yung,
{\em Non-Abelian vortices in ${\cal N} = 1^*$ gauge theory,}
hep-th/0408235.
%%CITATION = HEP-TH 0408235;%%

\bibitem{Armoni-two}
A.~Armoni and M.~Shifman,
%``Remarks on stable and quasi-stable k-strings at large N,''
Nucl.\ Phys.\ B {\bf 671}, 67 (2003)
[hep-th/0307020].
%%CITATION = HEP-TH 0307020;%%

\bibitem{oldstrings}
H.~J.~de Vega and F.~A.~Schaposnik,
 %``Electrically Charged Vortices In Non-Abelian
 %Gauge Theories With Chern-Simons
%Term,''
Phys.\ Rev.\ Lett.\  {\bf 56}, 2564 (1986);
%%CITATION = PRLTA,56,2564;%%
%``Vortices And Electrically Charged Vortices
%In Non-Abelian Gauge Theories,''
Phys.\ Rev.\ D {\bf 34}, 3206 (1986);\\
%%CITATION = PHRVA,D34,3206;%%
%\bibitem{HV}
J.~Heo and T.~Vachaspati,
%``Z(3) strings and their interactions,''
Phys.\ Rev.\ D {\bf 58}, 065011 (1998)
[hep-ph/9801455];\\
%%CITATION = HEP-PH 9801455;%%
%\bibitem{Su}
P.~Suranyi,
%``Vortex solutions in SU(N) adjoint Higgs theories,''
Phys.\ Lett.\ B {\bf 481}, 136 (2000)
[hep-lat/9912023];\\
%%CITATION = HEP-LAT 9912023;%%
%\bibitem{SS}
F.~A.~Schaposnik and P.~Suranyi,
%``New vortex solution in SU(3) gauge-Higgs theory,''
Phys.\ Rev.\ D {\bf 62}, 125002 (2000)
[hep-th/0005109];\\
%%CITATION = HEP-TH 0005109;%%
%\bibitem{KB}
M.~A.~C.~Kneipp and P.~Brockill,
%``BPS string solutions in non-Abelian Yang--Mills theories,''
Phys.\ Rev.\ D {\bf 64}, 125012 (2001)
[hep-th/0104171];\\
%%CITATION = HEP-TH 0104171;%%
%\bibitem{KoS}
K.~Konishi and L.~Spanu,
%``Non-Abelian vortex and confinement,''
Int.\ J.\ Mod.\ Phys.\ A {\bf 18}, 249 (2003)
[hep-th/0106175].
%%CITATION = HEP-TH 0106175;%%

\bibitem{Dorey:1998yh}
N.~Dorey,
%``The BPS spectra of two-dimensional supersymmetric gauge 
%theories with twisted mass terms,''
JHEP {\bf 9811}, 005 (1998)
[hep-th/9806056].

\bibitem{Dorey:1999zk}
N.~Dorey, T.~J.~Hollowood and D.~Tong,
%``The BPS spectra of gauge theories in two and four dimensions,''
JHEP {\bf 9905}, 006 (1999)
[hep-th/9902134].

\bibitem{VY}
A.~I.~Vainshtein and A.~Yung,
%``Type I superconductivity upon
%monopole condensation in Seiberg-Witten  theory,''
Nucl.\ Phys.\ B {\bf 614}, 3 (2001)
[hep-th/0012250].
%%CITATION = HEP-TH 0012250;%%

\bibitem{BarH}
K.~Bardakci and M.~B.~Halpern,
%``Spontaneous Breakdown And Hadronic Symmetries,''
Phys.\ Rev.\ D {\bf 6}, 696 (1972).
%%CITATION = PHRVA,D6,696;%%

\bibitem{MY}
A.~Marshakov and A.~Yung,
%``Non-Abelian confinement via Abelian
%flux tubes in softly broken N = 2  SUSY QCD,''
Nucl.\ Phys.\ B {\bf 647}, 3 (2002)
[hep-th/0202172].
%%CITATION = HEP-TH 0202172;%%

\bibitem{haho}
A.~Hanany and K.~Hori,
%``Branes and N = 2 theories in two dimensions,''
Nucl.\ Phys.\ B {\bf 513}, 119 (1998)
[hep-th/9707192].

\bibitem{NSVZsigma}
V.~A.~Novikov, M.~A.~Shifman, A.~I.~Vainshtein and V.~I.~Zakharov,
%``Two-Dimensional Sigma Models: Modeling 
%Nonperturbative Effects Of Quantum
%Chromodynamics,''
Phys.\ Rept.\  {\bf 116}, 103 (1984).
%%CITATION = PRPLC,116,103;%%

\bibitem{5}
E.~Witten,
%``Instantons, The Quark Model, And The 1/N Expansion,''
Nucl.\ Phys.\ B {\bf 149}, 285 (1979).
%%CITATION = NUPHA,B149,285;%%
\bibitem{da}
A.~D'Adda, M.~L\"{u}scher and P.~Di Vecchia,
%``A 1/N Expandable Series Of Nonlinear Sigma Models With Instantons,''
Nucl.\ Phys.\ B {\bf 146}, 63 (1978).
%%CITATION = NUPHA,B146,63;%%
\bibitem{nvacym}
E.~Witten,
%``Theta dependence in the large N limit of 
%four-dimensional gauge  theories,''
Phys.\ Rev.\ Lett.\  {\bf 81}, 2862 (1998)
[hep-th/9807109].
%%CITATION = HEP-TH 9807109;%%

\bibitem{nvacymp}
M.~Shifman,
%``Domain walls and decay rate of the 
%excited vacua in the large N  Yang--Mills
%theory,''
Phys.\ Rev.\ D {\bf 59}, 021501 (1999)
[hep-th/9809184].
%%CITATION = HEP-TH 9809184;%%

\bibitem{Acharya}
B.~S.~Acharya and C.~Vafa,
{\em On domain walls of ${\cal N} = 1$ supersymmetric Yang--Mills 
in four dimensions,} hep-th/0103011.

\bibitem{Volo}
I.~Y.~Kobzarev, L.~B.~Okun and M.~B.~Voloshin,
%``Bubbles In Metastable Vacuum,''
Yad.\ Fiz.\  {\bf 20}, 1229 (1974)
[Sov.\ J.\ Nucl.\ Phys.\  {\bf 20}, 644 (1975)];
%%CITATION = SJNCA,20,644;%%
M.~B.~Voloshin,
{\em False vacuum decay,}
 Lecture given at International School of Subnuclear Physics: 
{\sl Vacuum and Vacua: the Physics of Nothing}, Erice, Italy, 
July 1995, 
in {\sl Vacuum and vacua: the physics of nothing}, Ed. A.~Zichichi 
(World Scientific, Singapore, 1996), p. 88--124. 

\bibitem{Seiberg}
N.~Seiberg,
%``Theta Physics At Strong Coupling,''
Phys.\ Rev.\ Lett.\  {\bf 53}, 637 (1984).
%%CITATION = PRLTA,53,637;%%

\bibitem{CecV}
S.~Cecotti and C.~Vafa,
%``On classification of N=2 supersymmetric theories,''
Commun.\ Math.\ Phys.\  {\bf 158}, 569 (1993)
[hep-th/9211097].
%%CITATION = HEP-TH 9211097;%%

\bibitem{Bais}
F.~A.~Bais,
%``The Topology Of Monopoles Crossing A Phase Boundary,''
Phys.\ Lett.\ B {\bf 98}, 437 (1981).
%%CITATION = PHLTA,B98,437;%%

\bibitem{PrVi}
J.~Preskill and A.~Vilenkin,
%``Decay of metastable topological defects,''
Phys.\ Rev.\ D {\bf 47}, 2324 (1993)
[hep-ph/9209210].
%%CITATION = HEP-PH 9209210;%%

\bibitem{Kn}
M.~A.~C.~Kneipp,
%``Z(k) string fluxes and monopole confinement in non-Abelian
%theories,''
Phys.\ Rev.\ D {\bf 68}, 045009 (2003)
[hep-th/0211049];
%%CITATION = HEP-TH 0211049;%%
%``Color superconductivity, Z(N) flux tubes and monopole confinement in
%deformed
%N = 2* super Yang--Mills theories,''
Phys.\ Rev.\ D {\bf 69}, 045007 (2004)
[hep-th/0308086].
%%CITATION = HEP-TH 0308086;%%

\bibitem{ABEK}
R.~Auzzi, S.~Bolognesi, J.~Evslin and  K.~Konishi,
%{\em Non-Abelian monopoles and vortices that confine them,}
Nucl. Phys. {\bf B686}, 119 (2004)
[hep-th/0312233];  \\
R.~Auzzi, S.~Bolognesi, J.~Evslin, K.~Konishi and H.~Murayama,
{\em Nonabelian monopoles}, hep-th/0405070;\\
R.~Auzzi, S.~Bolognesi and  J.~Evslin,
{\em Monopoles can be confined by 0, 1 or 2 vortices}, hep-th/0411074.


\bibitem{INOS}
Y.~Isozumi, M.~Nitta, K.~Ohashi and N.~Sakai,
{\em All Exact Solutions of a 1/4 Bogomol'nyi-Prasad-Sommerfield 
Equation,} hep-th/0405129.
%%CITATION = HEP-TH 0405129;%%



\bibitem{W93}
E.~Witten,
%''Phases of \ntwo theories in two dimensions,''
Nucl.\ Phys.\ B {\bf 403}, 159 (1993)
[hep-th/9301042].

\bibitem{gvy}
A.~Gorsky, A.~I.~Vainshtein and A.~Yung,
%``Deconfinement at the Argyres-Douglas point in SU(2) gauge theory with
%broken N = 2 supersymmetry,''
Nucl.\ Phys.\ B {\bf 584}, 197 (2000)
[hep-th/0004087].

\bibitem{W97}
E.~Witten,
%``On the conformal field theory of the Higgs branch,''
JHEP {\bf 9707}, 003 (1997)
[hep-th/9707093].
%%CITATION = HEP-TH 9707093;%%

\bibitem{Ooguri}
H.~Ooguri and C.~Vafa,
%``Worldsheet derivation of a large N duality,''
Nucl.\ Phys.\ B {\bf 641}, 3 (2002)
[hep-th/0205297].
%%CITATION = HEP-TH 0205297;%%

\bibitem{SBone}
A.~Zamolodchikov and  Al.~Zamolodchikov,
Ann.~Phys. {\bf 120} (1979) 253.

\bibitem{Zam}
A.~B.~Zamolodchikov and A.~B.~Zamolodchikov,
%``Massless factorized scattering and 
%sigma models with topological terms,''
Nucl.\ Phys.\ B {\bf 379} (1992) 602.

\bibitem{Fat}
V.~A.~Fateev, E.~Onofri and A.~B.~Zamolodchikov,
%``The Sausage model (integrable deformations of O(3) sigma model),''
Nucl.\ Phys.\ B {\bf 406} (1993) 521.

\bibitem{Affl}
I.~Affleck and F.~D.~M.~Haldane,
%Critical theory of quantum chains,
Phys.\ Rev.\ B {\bf 36}, 5291 (1987);\\
I.~Affleck,
% ``Nonlinear Sigma Model At Theta = 
%Pi: Euclidean Lattice Formulation And
%Solid-On-Solid Models,''
Phys.\ Rev.\ Lett.\  {\bf 66}, 2429 (1991).

\bibitem{Cont}
D.~Controzzi and G.~Mussardo,
%``On the mass spectrum of the two-dimensional 
%O(3) sigma model with theta term,''
Phys.\ Rev.\ Lett.\  {\bf 92}, 021601 (2004)
[hep-th/0307143].
%%CITATION = HEP-TH 0307143;%%

\bibitem{Coleman}
S.~R.~Coleman,
%``More About The Massive Schwinger Model,''
Annals Phys.\  {\bf 101}, 239 (1976).


\end{thebibliography}
\end{document}